\documentclass[twocolumn]{aastex631}
\usepackage{amsmath}
\usepackage{color}

\shorttitle{Criteria for Dynamical MT}
\shortauthors{Ge et al.}

\begin{document}

\title{Criteria for Dynamical Timescale Mass Transfer of Metal-poor Intermediate-mass Stars}
\author[0000-0002-6398-0195]{Hongwei Ge}
\affiliation{Yunnan Observatories, Chinese Academy of Sciences,\\
396 YangFangWang, Guandu District, Kunming, 650216, People's Republic of China}
\affiliation{Institute of Astronomy, The Observatories, University of Cambridge,\\
	Madingley Road, Cambridge, CB3 0HA, UK}
\affiliation{Key Laboratory for Structure and Evolution of Celestial Objects, \\
	Chinese Academy of Sciences, P.O. Box 110, Kunming 650216, People's Republic of China}
\email{gehw@ynao.ac.cn}

\author[0000-0002-1556-9449]{Christopher A Tout}
\affiliation{Institute of Astronomy, The Observatories, University of Cambridge,\\
	Madingley Road, Cambridge, CB3 0HA, UK}
\email{cat@ast.cam.ac.uk}


\author[0000-0001-5284-8001]{Xuefei Chen}
\affiliation{Yunnan Observatories, Chinese Academy of Sciences,\\
	396 YangFangWang, Guandu District, Kunming, 650216, People's Republic of China}
\affiliation{Key Laboratory for Structure and Evolution of Celestial Objects, \\
	Chinese Academy of Sciences, P.O. Box 110, Kunming 650216, People's Republic of China}
\email{cxf@ynao.ac.cn}

\author[0000-0002-1455-2784]{Arnab Sarkar}
\affiliation{Institute of Astronomy, The Observatories, University of Cambridge,\\
	Madingley Road, Cambridge, CB3 0HA, UK}
\email{as3158@cam.ac.uk}

\author[0000-0001-5819-3552]{Dominic J Walton}
\affiliation{Centre for Astrophysics Research, University of Hertfordshire, College Lane, Hatfield AL10 9AB, UK}
\email{dwalton354@gmail.com}

\author[0000-0001-9204-7778]{Zhanwen Han}
\affiliation{Yunnan Observatories, Chinese Academy of Sciences,\\
396 YangFangWang, Guandu District, Kunming, 650216, People's Republic of China}
\affiliation{Key Laboratory for Structure and Evolution of Celestial Objects, \\
	Chinese Academy of Sciences, P.O. Box 110, Kunming 650216, People's Republic of China}
\affiliation{University of Chinese Academy of Sciences, Beijing 100049, People's Republic of China}
\email{zhanwenhan@ynao.ac.cn}

\begin{abstract}
The stability criteria of rapid mass transfer and common envelope evolution are fundamental in binary star evolution. They determine the mass, mass ratio and orbital distribution of many important systems, such as X-ray binaries, Type Ia supernovae and merging gravitational wave sources. We use our adiabatic mass-loss model to systematically survey the intermediate-mass stars' thresholds for dynamical-timescale mass transfer. The impact of metallicity on the stellar responses and critical mass ratios is explored. Both tables ($Z=0.001$) and fitting formula ($Z=0.001$ and $Z=0.02$) of critical mass ratios of intermediate-mass stars are provided. An application of our results to intermediate-mass X-ray binaries (IMXBs) is discussed. We find that the predicted upper limit to mass ratios, as a function of orbital period, is consistent with the observed IMXBs that undergo thermal or nuclear timescale mass transfer. According to the observed peak X-ray luminosity $L_\mathrm{X}$, we predict the range of $L_\mathrm{X}$ for IMXBs as a function of the donor mass and the mass transfer timescale.

\end{abstract}


\section{Introduction} \label{sec:1}

The fraction of binary including multiple stars is over half of stellar systems \citep[e.g.][]{2013ARA&A..51..269D,2017ApJS..230...15M,2022ApJ...933..119L} and it can be even up to $70\%$ for massive stars \citep{2012Sci...337..444S}. The evolution of close binary stars can form X-ray binaries, pulsar binaries, type Ia supernovae, white dwarf/neutron star (NS)/stellar-mass black hole (BH) binaries, etc. The stability of rapid mass transfer and the common envelope evolution \citep{1976IAUS...73...75P} are fundamental problems in binary evolution and determine the fate of binary systems. Recent studies \citep{2015ApJ...812...40G,2020ApJ...899..132G,2017MNRAS.465.2092P,2021A&A...650A.107M,2023A&A...669A..45T} suggest that the critical initial mass ratios for dynamical-timescale mass transfer are larger than previously expected from polytropic stellar models, with the exception of massive early main-sequence (MS) stars \citep{2015ApJ...812...40G,2020ApJ...899..132G}. So, binaries in a stable mass transfer channel contribute significantly to merging BHs \citep[e.g.][]{2017MNRAS.468.5020I,2021ApJ...922..110G,2022arXiv220613842B,2022arXiv220708837D}.

Specifically, intermediate-mass X-ray binaries (IMXBs) are important and energetic objects among binary systems with donor mass $1.5 < M/M_\odot < 10.0$. They are rare and little studied previously compared with low- and high-mass X-ray binaries. However, for the current low-mass X-ray binary (LMXB) Cygnus X-2, \citet{1999MNRAS.309..253K} and \citet{2000ApJ...529..946P} independently suggest the luminous and hot companion ($M \approx 0.5 M_\odot$) formed through non-conserved and super-Eddington thermal timescale mass transfer from a previous $M \approx 3.5 M_\odot$ star. \citet{2000ApJ...530L..93T} provide a detailed calculation of IMXBs with $2$-$6\,M\odot$ donor and $1.3\,M\odot$ accretor and demonstrate that in many cases that systems will evolve to binary millisecond pulsars. \citet{2002ApJ...565.1107P} present systematically the evolution of I/LMXBs with $0.6$-$7\,M\odot$ donors. \citet{2012ApJ...756...85S} further present a systematic study of I/LMXBs with different neutron star masses. \citet{2020AA...642A.174M} show that observed super-Eddington luminosities can be achieved in I/LMXBs undergoing a non-conserved mass transfer. Clearly, the upper limit of the initial mass ratio ($q=M_\mathrm{donor}/M_\mathrm{accretor}$) to form I/LMXBs should be, in principle, consistent with the critical initial mass ratio for dynamical timescale mass transfer. The allowed parameter space of the initial orbital period and donor mass from the above studies suggests the critical initial mass ratio $q \approxeq 3-4$ for dynamical timescale mass transfer of radiative donor stars. This is in agreement with studies by \citet{1989PhDT.........7H} and \citet{1996ApJ...458..301K} and is widely adopted in binary population synthesis codes for radiative MS/Hertzsprung gap (HG) donor stars.

After the mass of a star, which is the most fundamental parameter, metallicity is the next most important parameter in stellar evolution. Many observed stellar phenomena including binaries are dominated by metal-poor environments. Examples include horizontal-branch stars \citep{1970ApJ...161..587I}, blue stragglers (blue metal-poor stars; \citealt{2000AJ....120.1014P}), Galactic halo stars \citep[e.g.][]{2006ChJAA...6..265Z,2018ApJS..238...16L}, metal-poor thick disk stars \citep[e.g.][]{2021MNRAS.501.4917W} and the stellar initial mass function of ultra-faint dwarf satellite galaxies \citep{2020A&A...637A..68Y}. Inspired by the important contribution to the chemical evolution of galaxies, asymptotic giant branch nucleosynthesis and supernovae physics, the abundances of C \citep{1974ApJ...189..493S}, N \citep{1973ApJ...184..839S}, O \citep{2004A&A...414..931A} and supernova elements \citep{1995AJ....109.2736M,1996ApJ...471..254R} are all affected.

\citet{2009MNRAS.400..677Z} suggest that super-Eddington accretion onto stellar-mass BHs at low-metallicity, rather than intermediate-mass BHs, contribute significantly to ultra-luminous X-ray sources (ULXs). \citet{2010ApJ...715L.138B} find that
the gravitational wave detection rate is increased by a factor of 20 if the metallicity is decreased from solar to a half-half mixture of solar and $10\%$ solar metallicity. The chemically homogeneous evolution of stars favours a low-metallicity environment \citep{2005A&A...443..643Y}. In addition to isolated binary evolution and dynamical interaction in a dense cluster, chemically homogeneous evolution of binaries is an important source of merging BHs \citep{2016MNRAS.458.2634M,2016MNRAS.460.3545D}. Gravitational wave detection discoveries are frequently merging massive stellar-mass BHs ($M>30M_\sun$) \citep{2021PhRvX..11b1053A}, which it has been suggested form in metal-poor environments \citep[e.g.][]{2021MNRAS.504..146V}. \citet{2020A&A...638A..55K} show that metallicity has a strong influence on the type of mass transfer in massive binary systems. \citet{2022A&A...662A..56K} find that the metallicity of massive stars strongly influences the course and outcome of mass-transfer evolution.

Here we focus on intermediate-mass (IM) stars ($1.6 \le M/M_\sun \le 10$) with metallicity $Z=0.001$. We make a comparison of the radius response of IM stars with different metallicites undergoing adiabatic mass loss. We find their critical mass ratios for dynamical-timescale mass transfer. An application to IMXBs is also presented. We briefly mention methods and stellar model selection in Section \ref{sec:met}. Using $4\,M_{\odot}$ stars as examples in Section \ref{sec:4m} we study the effect of metallicity on the response of stars to adiabatic mass loss. We provide the critical mass ratios for dynamical-timescale mass transfer of the IM stars in Section \ref{sec:res}. Fitting formula for these critical mass ratios for both $Z=0.001$ and $Z=0.02$ IM stars are provided in Section \ref{sec:fit}. In Section \ref{sec:imxbs} and \ref{sec:sum}, we apply our results to observed IMXBs and summarize our studies respectively.

\section{Methods and model selections}
\label{sec:met}

\begin{figure}[ht!]
	\centering
	\includegraphics[scale=0.32]{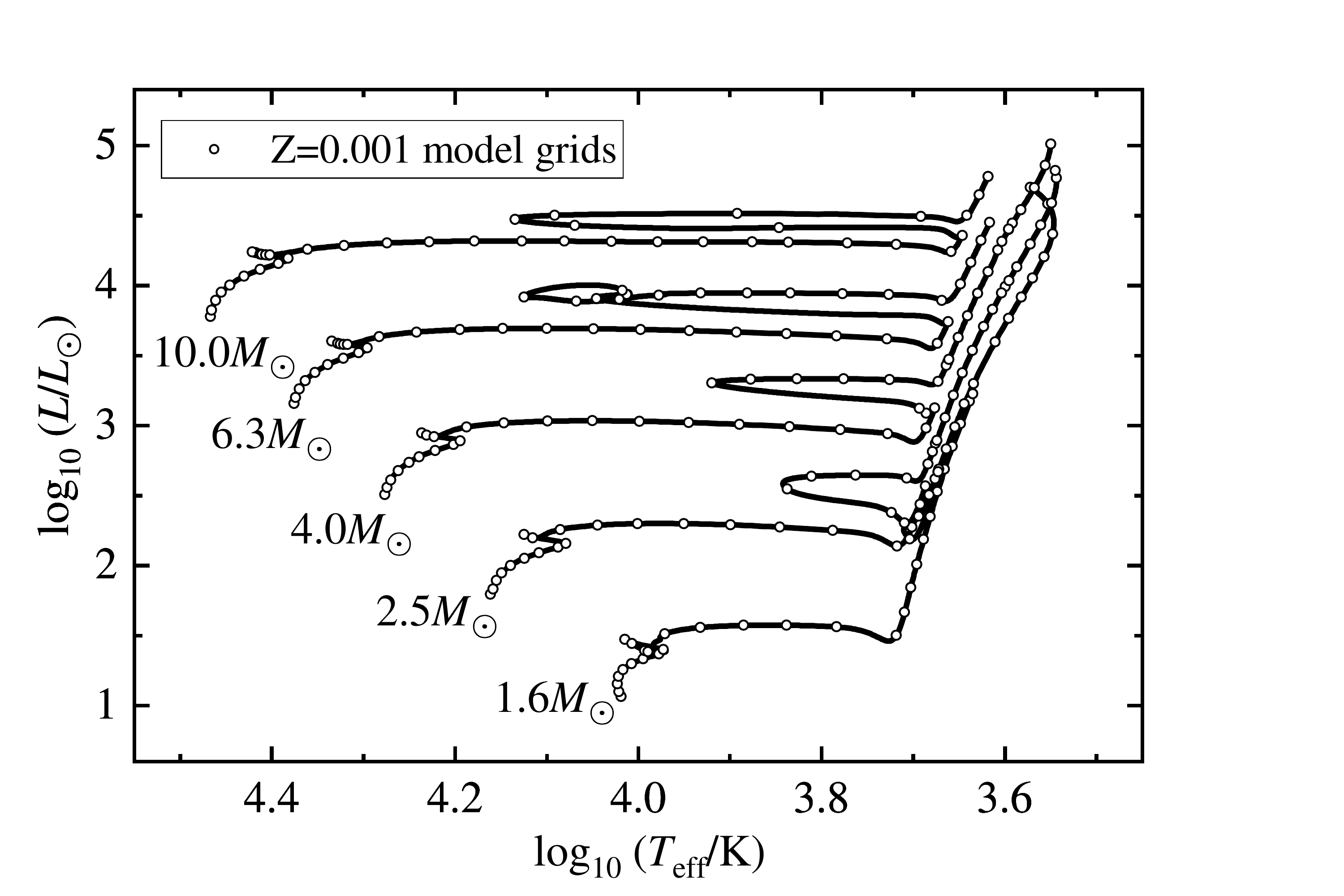}
	\caption{Hertzsprung-Russell diagram of intermediate-mass (IM) stars and model grids (circles). The masses are labelled and the metallicity is $Z=0.001$.}
	\label{grids}
\end{figure}

We use our adiabatic mass-loss model to study the responses of IM donor stars with metallicity $Z=0.001$. Methods and numerical implementations are described in detail in Papers I, II and III \citep{2010ApJ...717..724G,2015ApJ...812...40G,2020ApJ...899..132G}. We use the same physical parameters, such as mixing-length and overshooting coefficients, for metal-poor IM stars. As for Papers I-III, we build initial model sequences that undergo adiabatic mass loss without including stellar winds. The masses of the initial models are 1.6, 2.5, 4.0, 6.3 and $10.0\,M_\sun$. Radius grids are selected roughly with $\Delta \log_{10} (R/R_\sun)=0.1$ except for the MS stars (Figure\,\ref{grids}). 

We introduce key points of caculating the critical mass ratio $q_\mathrm{ad}$ for dynamical timescale mass transfer. In principle, the critical initial mass ratio is the minimum value satisfied the mass-radius exponent of the donor star $\zeta_\mathrm{ad} = \mathrm{d} \ln R/\mathrm{d} \ln M$ which equal to the mass-radius exponent of its Roche lobe $\zeta_\mathrm{L} (q_\mathrm{ad})=\mathrm{d} \ln R_\mathrm{L}/\mathrm{d} \ln M$ throughout the whole adiabatic mass loss process. Because the runaway mass trasfer is increased gradually as mass transfer begins. So, instead of using the surface radius of the donor star $R$, we use an inner radius $R_\mathrm{KH}$ to calculate the mass-radius exponent $\zeta_\mathrm{KH} = \mathrm{d} \ln R_\mathrm{KH}/\mathrm{d} \ln M |_\mathrm{ad}$. This innner radius represents the mass loss rate $\dot{M}$ (see A9 in \citealt{2010ApJ...717..724G}) reaching a thermal timescale rate $\dot{M}_\mathrm{KH} = M_\mathrm{i}/\tau_\mathrm{KH}$.

We count the model number $n$ from 1 to $\mathrm{N}$ for the whole adiabatic mass loss process. For the initial model $n=1$, we have the initial mass $M_\mathrm{1i}$ and initial radius $R_\mathrm{1i}$. The Roche-lobe radius of this model is $R_\mathrm{L,i} = R_\mathrm{1i}$. The initial mass ratio $q_\mathrm{i} = M_\mathrm{1i}/M_\mathrm{2i}$ is unknown and to be solved. We define a mass function $\mu = M_1/(M_1 + M_2) = q/(1+q)$ for convenience since it can only change from 0 to 1. For model number $n$, the mass $M_n$ and inner radius $R_{\mathrm{KH},n}$ are solved from the adiabatic mass loss calculation. The mass-radius exponent $\zeta_\mathrm{ad} = \zeta_\mathrm{KH}$ can be calculated from models $n$ and $n-1$. If we assume the mass transfer is conserved in mass and angular momentum, the mass and Roche-lobe radius exponent $\zeta_\mathrm{L}$ is a function of the mass ratio (see Equation 45 in \citealt{2010ApJ...717..724G}). Applying to the orbital angular momentum of binary, for conserved mass transfer, we can write 
\begin{equation}
\frac{R_\mathrm{L}(\mu_n)}{R_\mathrm{1i}} = \frac{r_\mathrm{L}(\mu_n)}{r_\mathrm{L}(\mu_\mathrm{i})} \left( \frac{\mu_\mathrm{i}}{\mu_n}\right)^2 \left( \frac{1-\mu_\mathrm{i}}{1-\mu_n} \right)^2,
\end{equation} 
where $\mu_n = \mu_\mathrm{i} M_n/M_\mathrm{1i}$, $\mu_\mathrm{i} = q_\mathrm{i}/(1+q_\mathrm{i})$, and
\begin{equation}
r_\mathrm{L}(q) = \frac{0.49 q^{2/3}}{0.6q^{2/3}+ \ln (1+q^{1/3})},
\end{equation}
from Eggleton's approximation \citep{1983ApJ...268..368E}. Starting from an initial guess $\mu_\mathrm{i} = 0.5$, we use a Bisection method to calculate the initial mass function $\mu_n$ satisfying both $R_{\mathrm{L},n} = R_{\mathrm{KH},n}$ and $\zeta_{\mathrm{L},n} = \zeta_{\mathrm{KH},n}$. By tracing $n=1$ to $n=$N, we can get the minimum value of $\mu_\mathrm{min}$. So, the critical initial mass ratio is calculated finally with $q_\mathrm{ad} = \mu_\mathrm{min} /(1-\mu_\mathrm{min})$.

In addition to standard donor stars with mixing-length convective envelopes, we also build parallel donor star sequences with isentropic envelopes. In these stars, convective envelopes have been replaced by isentropic ones, with specific entropy fixed to be that at the base of envelopes. By doing this, the limitations of adiabatic approximation at the tiny layer under the photosphere are overcome. Consequently, the superadiabatic expansion in a donor star with a thick convective envelope becomes placid. We can find a more detailed explanation in the papers I, II, and III. The critical initial mass ratios $\tilde{q}_\mathrm{ad}$ for these donor stars can be calculated with the same method mentioned above. A $\sim$ script on the top of corresponding parameters, such as the inner radius $\tilde{R}_\mathrm{KH}$, is labeled for donor stars with replaced envelopes.

\section{$4\,M_{\odot}$ stars with different metallicity}
\label{sec:4m}

\begin{figure}[ht!]
\centering
\includegraphics[scale=0.29]{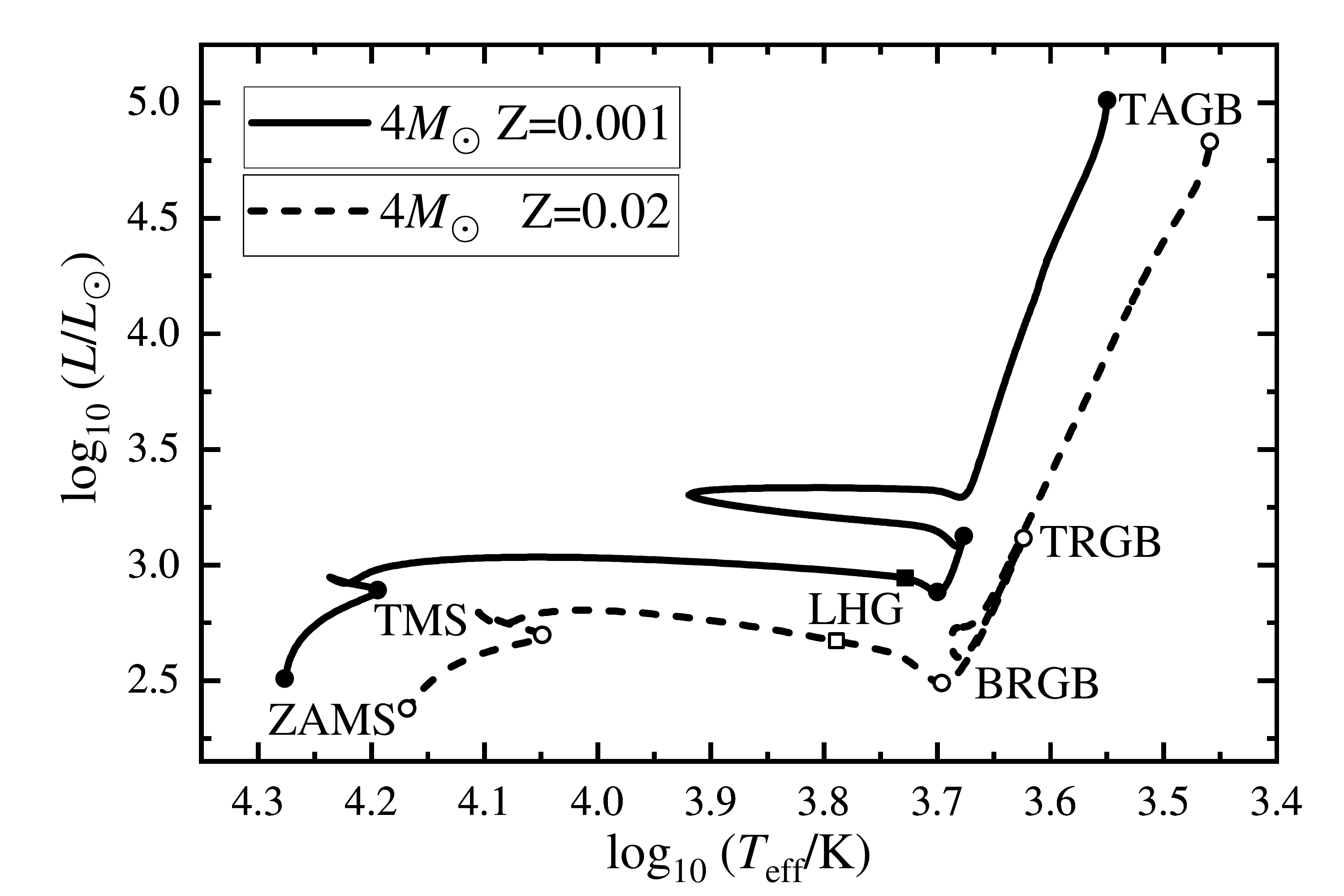}
\caption{Theoretical Hertzsprung-Russell diagram of $4\,M_{\odot}$ stars with metallicity $Z=0.001$ (solid line) and $Z=0.02$ (dashed line). Solid and open circles show the location of important evolutionary stages, such as the zero-age main sequence (ZAMS), the terminal of the main sequence (TMS), the base of the red giant branch (BRGB), the tip of the red giant branch (TRGB) and the tip of the asymptotic giant branch (TAGB). Filled and open squares are the late Hertzsprung-Russell gap (LHG) where the critical mass ratio reaches a maximum.}
\label{4m-hrd}
\end{figure}
\begin{figure}[ht!]
\centering
\includegraphics[scale=0.29]{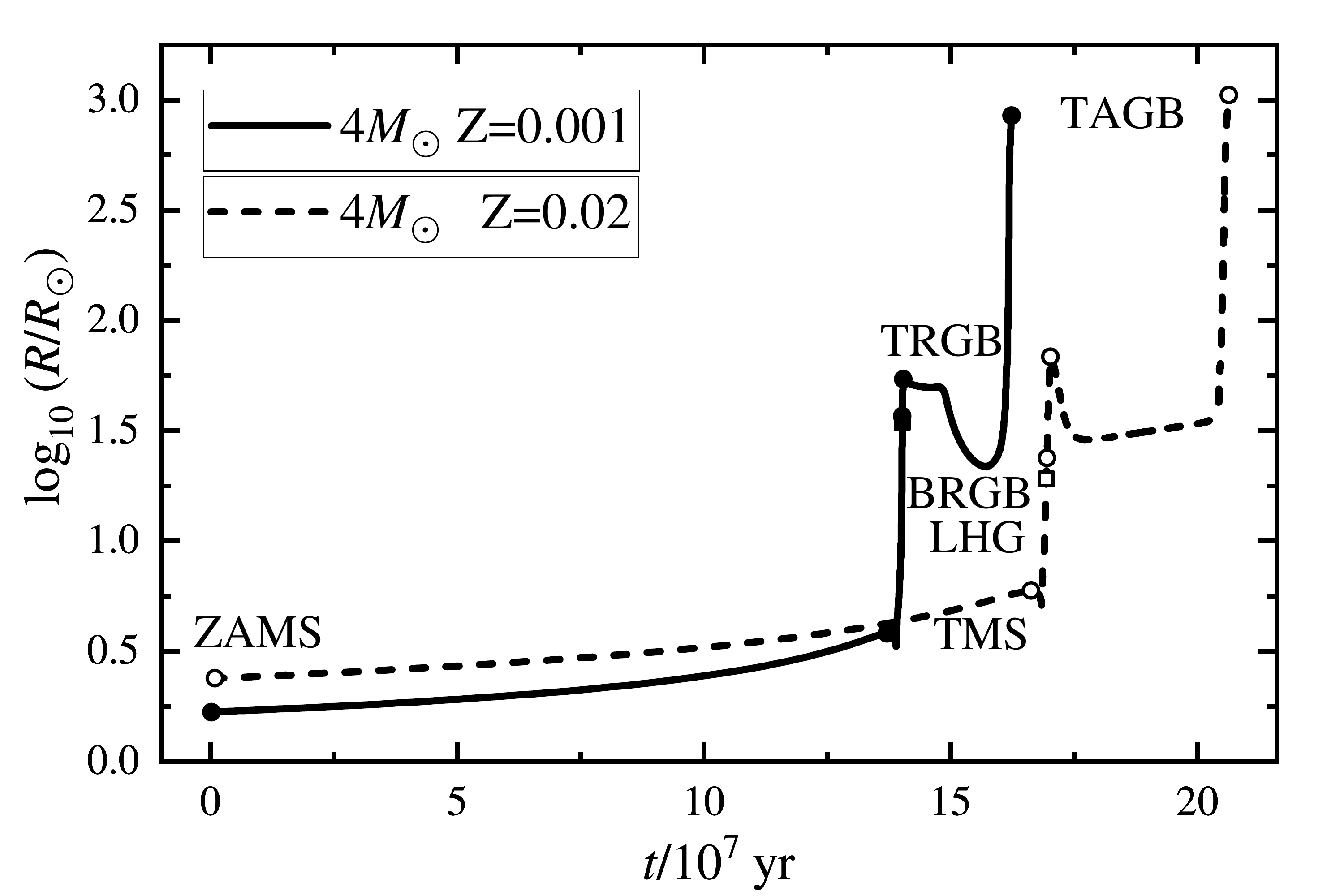}
\caption{Radii $R$ of $4\,M_{\odot}$ stars with metallicity $Z=0.001$ (solid line) and $Z=0.02$ (dashed line) as a function of age $t$. Lines and symbols correspond with that of Figure~\ref{4m-hrd}.}
\label{4m-rt}
\end{figure}

In this section, we study the impact of metallicity on the critical mass ratio $\tilde{q}_\mathrm{ad}$ for dynamical-timescale mass transfer of IM stars. To understand metallicity effects, we first consider the differences in the global physical behaviour of a $M_\mathrm{1i} = 4\,M_{\odot}$ star with metallicity $Z=0.001$ and $Z=0.02$. Secondly, using the terminal main-sequence (TMS) and the tip of the red giant branch (TRGB) models, we examine the response to adiabatic mass loss of $4\,M_{\odot}$ stars with $Z=0.001$ and $Z=0.02$ of different radii. Lastly, we calculate the difference in $\tilde{q}_\mathrm{ad}$ between solar metallicity and metal-poor $4\,M_{\odot}$ donor stars. We use $Z=0.02$ for solar metallicity despite recent studies indicating a lower metallicity in the solar atmosphere \citep[e.g.][]{2009ARA&A..47..481A}. 
\begin{figure}[ht!]
\centering
\includegraphics[scale=0.32]{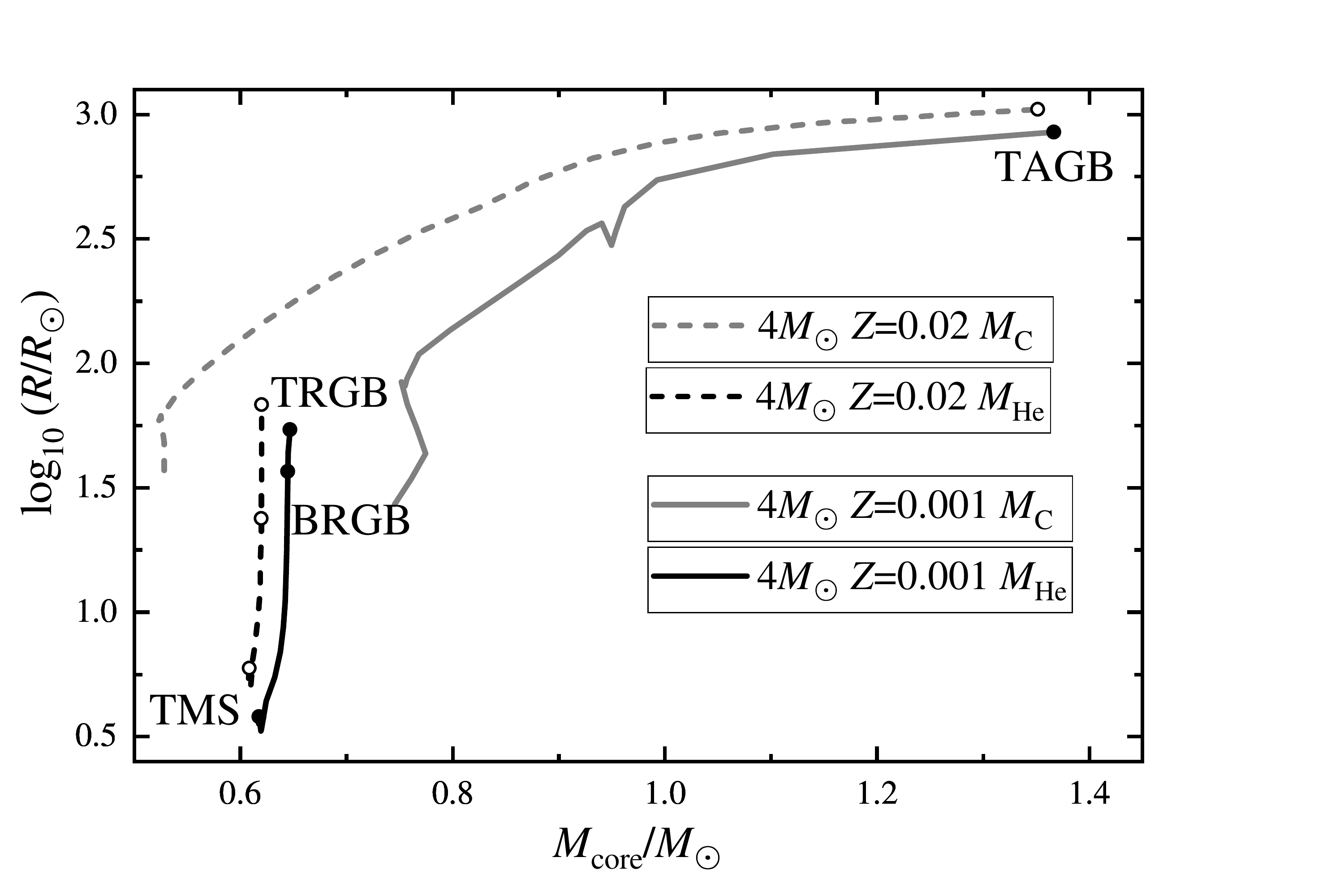}
\caption{Radii of $4\,M_{\odot}$ stars with metallicity $Z=0.001$ (solid lines) and $Z=0.02$ (dashed lines) as a function of their core masses. The helium core mass ($M_\mathrm{He}$ black lines) is where the maximum mass fraction of hydrogen is 0.15, and the carbon core mass ($M_\mathrm{C}$ gray lines) is where the maximum mass fraction of helium is 0.25. Solid and open circles correspond with those in Figure~\ref{4m-hrd}. Note that the helium core only appears after the hydrogen is exhausted in the convective core.}
\label{4m-rmc}
\end{figure}

It is well known that metal-poor MS stars are more compact, hotter and of smaller radii \citep[e.g.][]{pols2011stellar}. Figure~\ref{4m-hrd} shows that a $4\,M_{\odot}$ star with $Z=0.001$ is more luminous and hotter than its $Z=0.02$ counterpart at every evolutionary stage. The radius of a $4 M_{\odot}$ metal-poor star is almost always smaller than that at solar metallicity with the only exception near the base of the red giant branch (BRGB, see Figure~\ref{4m-rt}). This leads to a slightly larger HG for the metal-poor star. However, it has a larger core-mass (see Figure~\ref{4m-rmc}). This leads to a smaller evolutionary range on the red giant branch.

The radiative envelope dominates its radius response to the adiabatic mass loss for IM stars on the MS or in the HG. Therefore, an initial shrinkage of the radius is expected during the mass loss. However, after the IM star evolves to the red giant branch (RGB) or the asymptotic giant branch (AGB) the rapidly growing convective envelope dominates the radius response to adiabatic mass loss. Therefore, an initial radius expansion is followed for a RGB/AGB star during the mass loss. Because the responses of donor stars with the radiative and convective envelope are different, we choose two stellar models at the TMS and TRGB as examples. In the following, we first show the critical mass ratio of two example models with $Z=0.001$. Then, we demonstrate the impact of metallicity on the critical mass ratio.

\begin{figure}[ht!]
	\centering
	\includegraphics[scale=0.29]{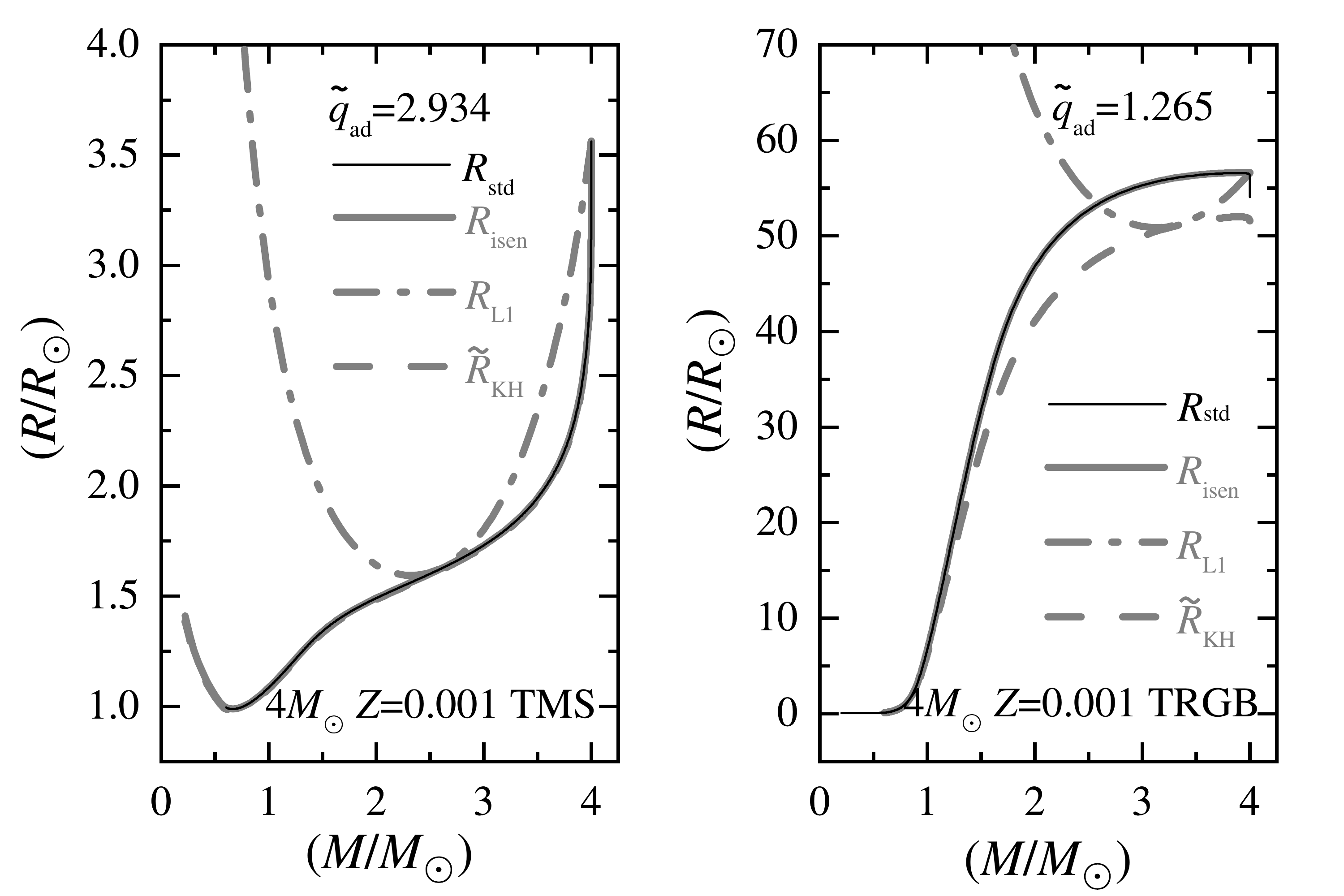}
	\caption{Radial response curves for a $4\,M_\sun$ ($Z=0.001$) TMS star (left panel) and a TRGB star. Solid lines trace the adiabatic mass-loss sequences. Thin black solid lines represent standard stars; thick gray solid lines represent correspondingly isentropic envelope stars. Gray dashed lines are shown for the inner radii $\tilde{R}_{\rm KH}$ at where the mass-loss rate reaches $\dot{M}_{\rm KH}$. Gray dash-dotted lines mark the Roche-lobe radius as a function of mass for critical initial mass ratios $\tilde{q}_\mathrm{ad}$. The corresponding limits for standard models such as $R_{\rm KH}$ and $q_\mathrm{ad}$ are omitted for clarity.}
	\label{4m-z001-q}
\end{figure}

We apply the calculation method described in the previous section to metal-poor $3.56\,R_\sun$ TMS and $54.09\,R_\sun$ TRGB models (Figure\,\ref{4m-z001-q}). The left panel of Figure\,\ref{4m-z001-q} shows the TMS donor's critical initial mass ratio $\tilde{q}_\mathrm{ad} = 2.934$. If the initial mass ratio $q_\mathrm{i} < \tilde{q}_\mathrm{ad}$ the mass transfer is dynamically stable and vice versa. The curves of the radius of the TMS donor star and its Roche-lobe radius show delayed dynamical instability. Its Roche-lobe radius curve tangent with the donor's inner radius at $\tilde{M}_{\rm KH} = 2.636\,M_\sun$. For this radiative envelope star, the inner radius $\tilde{R}_{\rm KH}$ and its isentropic envelope radius are almost identical to its radius during the adiabatic mass loss. The right panel of Figure\,\ref{4m-z001-q} presents the TRGB donor's critical initial mass ratio $\tilde{q}_\mathrm{ad} = 1.265$. The inner radius of this TRGB donor tangents with its Roche-lobe radius at $\tilde{M}_{\rm KH} = 3.328\,M_\sun$. This TRGB donor's extended and low-density convective envelope makes the inner radius $\tilde{R}_{\rm KH}$ much smaller than its radius but not the same after the core is exposed. We expect binary systems with this TRGB donor will evolve to the common envelope phase if the initial mass ratio is more significant than 1.265.

\begin{figure}[ht!]
	\centering
	\includegraphics[scale=0.32]{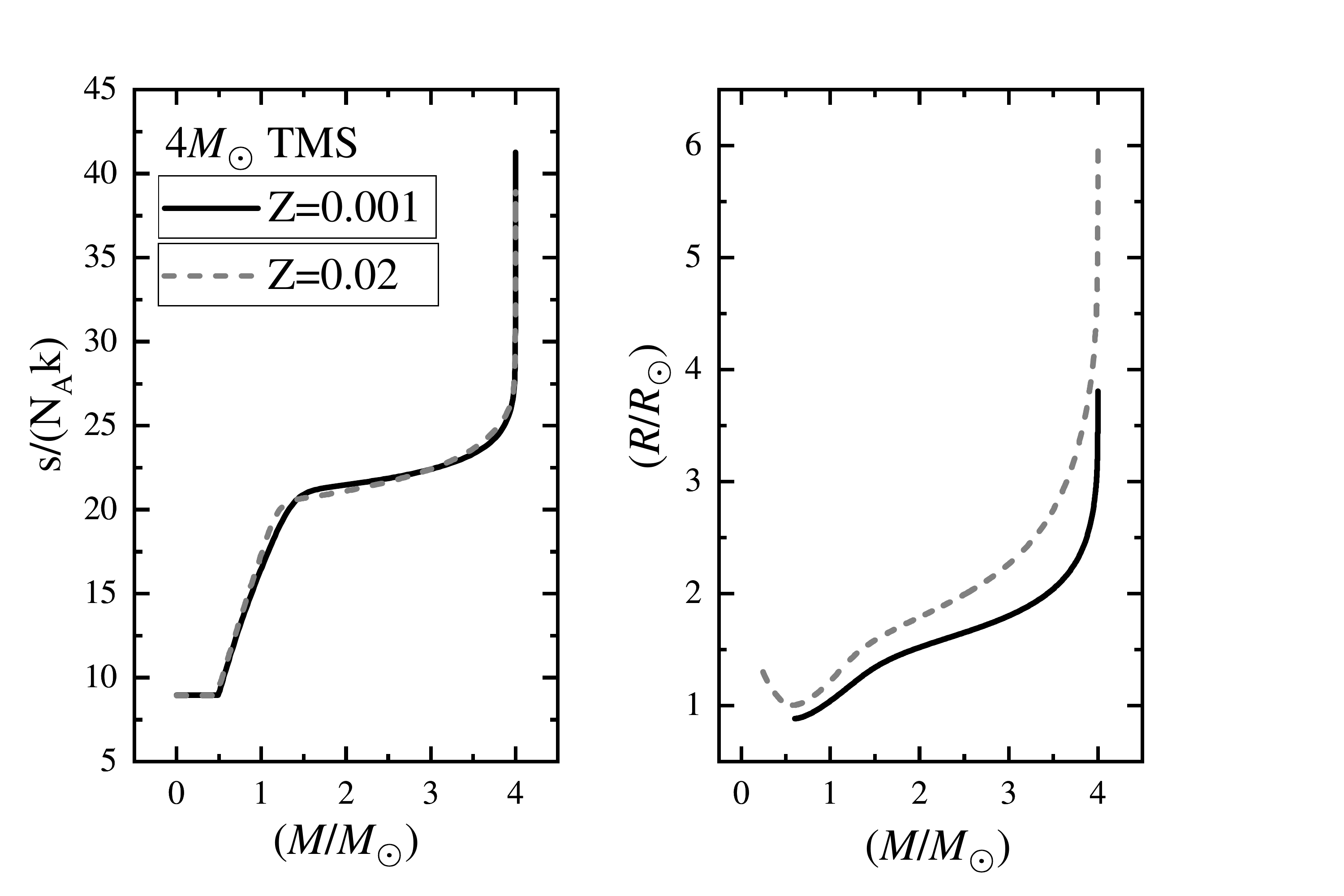}
	\caption{Specific entropy $s$ profile (left panel) and the remnant radius $R$ as a function of mass $M$ (right panel). Solid and dashed lines present $4 M_{\odot}$ TMS stars with $Z=0.001$ and $Z=0.02$. We used gray dashed lines to make the overlap region distinguishable.}
	\label{4m-tms}
\end{figure}
\begin{figure}[ht!]
	\centering
	\includegraphics[scale=0.29]{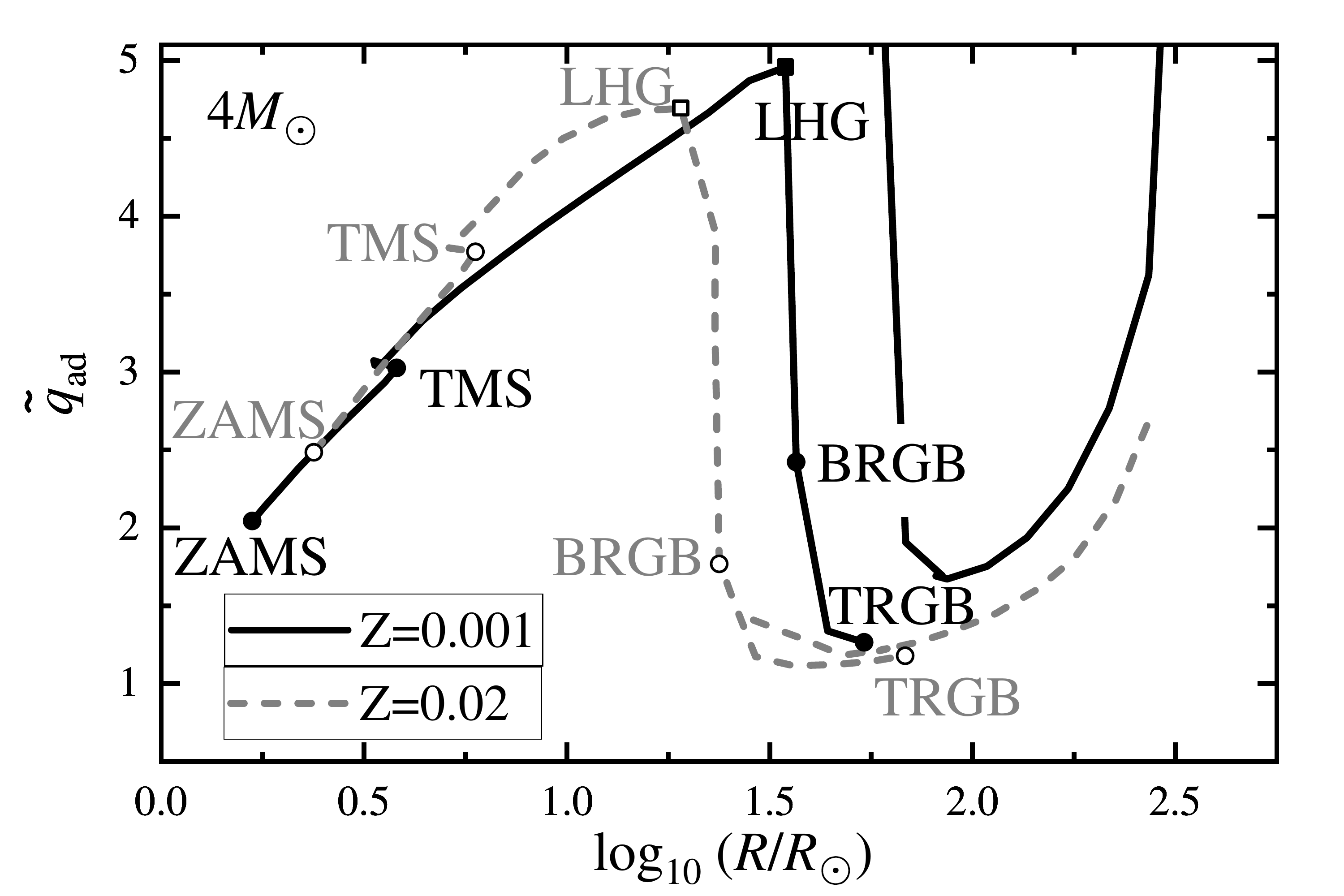}
	\caption{Critical mass ratios $\tilde{q}_\mathrm{ad}$ as a function of the donor's initial radius $R$. Lines and symbols have the same meaning as in Figures~\ref{4m-hrd} and \ref{4m-rt}.}
	\label{4m-zq}
\end{figure}

The differences in the entropy profile between two TMS stars with $Z=0.001$ and $Z=0.02$ are negligible. So the radius responses have the same trend (Figure~\ref{4m-tms}). However, the $Z=0.001$ TMS star has a larger convective core (Figure~\ref{4m-rmc}) and a smaller radiative envelope. This diminishes the contraction of the metal-poor star (see the right panel of Figure~\ref{4m-tms}). Consequently, the critical mass ratio for dynamical-timescale mass transfer of the metal-poor star at each evolutionary stage is smaller (Figure~\ref{4m-zq}). The critical mass ratios of donor stars with different metallicity have the same trend. $\tilde{q}_\mathrm{ad}$ depends on its evolutionary stage for a given mass star. From ZAMS to the late Hertzsprung-Russell gap (LHG; where $\tilde{q}_\mathrm{ad}$ reaches a maximum) $\tilde{q}_\mathrm{ad}$ increases almost linearly with the logarithm of the stellar radius. Then, a sudden drop of $\tilde{q}_\mathrm{ad}$ indicates the switching from a radiative-dominated structure to a convective-dominated one. From slightly late BRGB to TAGB (neglecting core-helium burning stages), $\tilde{q}_\mathrm{ad}$ increases again with the radius.
\begin{figure}[ht!]
	\centering
	\includegraphics[scale=0.32]{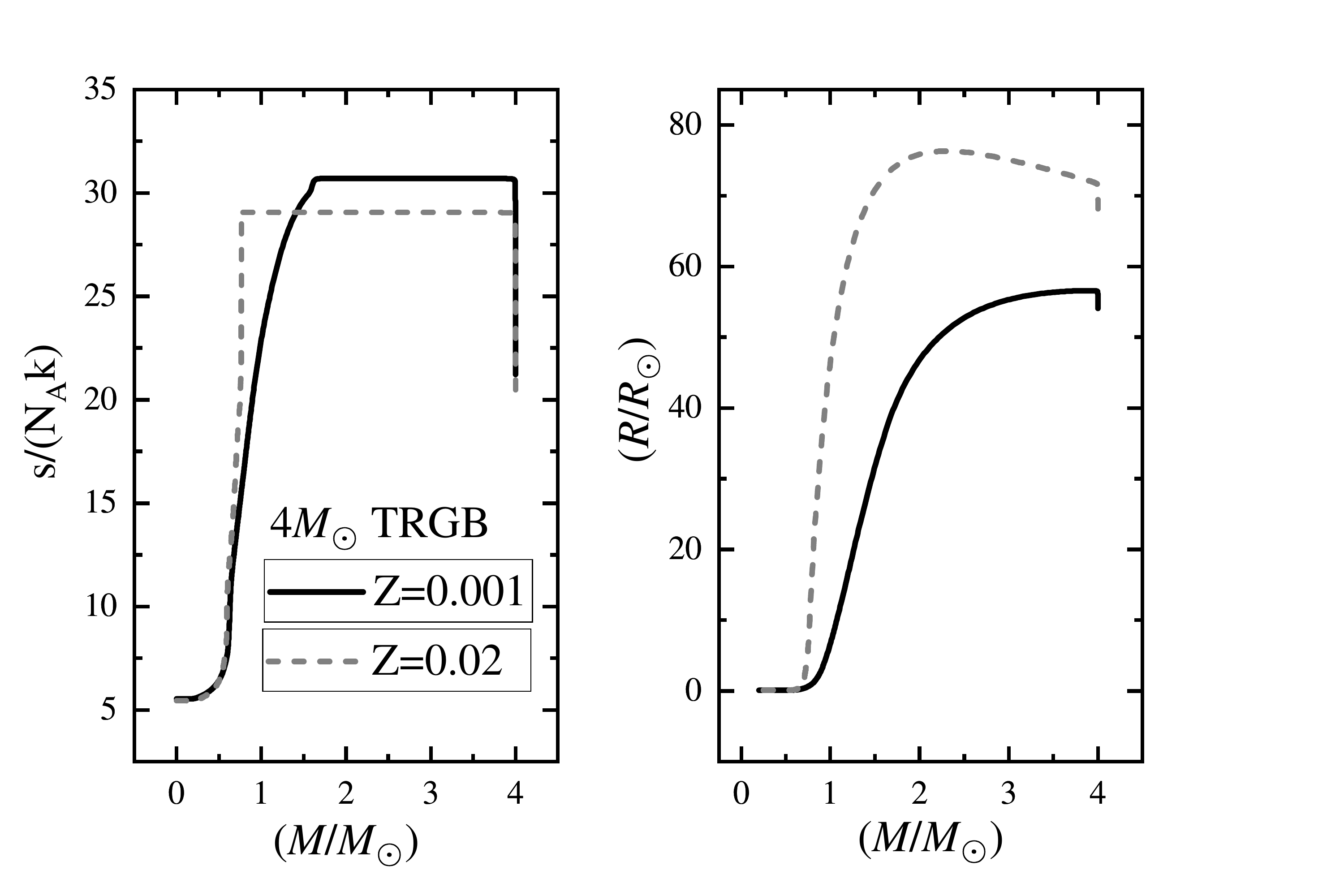}
	\caption{As Figure~\ref{4m-tms} but for $4 M_{\odot}$ TRGB stars.}
	\label{4m-trgb}
\end{figure}

At the TRGB the radius response is dominated by the deep convective envelope. But the partial ionization and non-ideal gas effects change the behaviour from that of the simplified polytropic models. In fact, critical mass ratios for dynamical-timescale mass transfer differ greatly between realistic stars and those with a polytropic equation of state (see Figure 9 in Paper III). The response of the thin layer under the photospheric surface might be dominated by radiation. The initial superadiabatic expansion in the right panel of Figure~\ref{4m-trgb} might be overestimated by the adiabatic assumption so we build isentropic envelope models to offset part of the superadiabatic expansion (see details in Papers I-III). The metal-poor model has a larger helium core than the solar metallicity model. So the convective envelope of the metal-poor star is thinner. Also the thermal timescale of metal-poor RGB/AGB stars is systematically shorter than that of the solar metallicity stars at the same radius. So, critical mass ratios of metal-poor stars are larger than the solar metallicity stars with the same radius.

In summary, for a metal-poor MS and HG donor star, we find $\tilde{q}_\mathrm{ad}$ is smaller than for a solar metallicity star at the same evolutionary stage. For a metal-poor RGB/AGB donor star we find $\tilde{q}_\mathrm{ad}$ is larger than for a solar metallicity star at the same radius.

\section{Results}
\label{sec:res}

\begin{deluxetable*}{rrrrrrrrrrrrrrrr}   
	\tabletypesize{\footnotesize}
	\tablewidth{0pt}
	\tablecolumns{16}
	\tablecaption{Properties and critical mass ratios of initial stellar models\label{4m-tab}}
	
	\tablehead{
		\colhead{$k$} & \colhead{$\log_{10} t$} & \colhead{$M$} & \colhead{$\log_{10} R$} &
		\colhead{$M_{\rm ce}$} & \colhead{$M_{\rm He}$} & \colhead{$M_{\rm C}$}
		&\colhead{$\log_{10} T_{\rm e}$} & \colhead{$\log_{10} L$} & \colhead{$X_{\rm s}$} &
		\colhead{$\rho_{\rm c}$} & \colhead{$T_{\rm c}$} & \colhead{$q_{\rm ad}$} &
		\colhead{$M_{\rm KH}$} & \colhead{$\tilde{q}_{\rm ad}$} & \colhead{$\tilde{M}_{\rm KH}$} \\
		\colhead{--} & \colhead{/yr} & \colhead{/$M_\sun$} & \colhead{/$R_\sun$} &
		\colhead{/$M_\sun$} & \colhead{/$M_\sun$} & \colhead{/$M_\sun$} &
		\colhead{/$\mathrm{K}$} & \colhead{/$L_\sun$} & \colhead{--} & 
		\colhead{/$\rm g\:cm^{-3}$} & \colhead{/$\mathrm{K}$} & \colhead{--} & \colhead{/$M_\sun$} & \colhead{--} & \colhead{/$M_\sun$} 
	}
	\startdata
1	&	--	&	4.0	&	0.224	&	0.000	&	0.000	&	0.000	&	4.277	&	2.508	&	0.756	&	1.613	&	7.462	&	2.042	&	2.478	&	2.044	&	2.477	\\*
2	&	7.4528	&	4.0	&	0.254	&	0.000	&	0.000	&	0.000	&	4.274	&	2.558	&	0.756	&	1.605	&	7.469	&	2.132	&	2.498	&	2.134	&	2.501	\\*
3	&	7.7343	&	4.0	&	0.289	&	0.000	&	0.000	&	0.000	&	4.270	&	2.612	&	0.756	&	1.603	&	7.478	&	2.237	&	2.525	&	2.238	&	2.525	\\*
4	&	7.9073	&	4.0	&	0.338	&	0.000	&	0.000	&	0.000	&	4.262	&	2.677	&	0.756	&	1.606	&	7.491	&	2.378	&	2.554	&	2.380	&	2.554	\\*
5	&	8.0020	&	4.0	&	0.391	&	0.000	&	0.000	&	0.000	&	4.250	&	2.736	&	0.756	&	1.616	&	7.505	&	2.523	&	2.581	&	2.525	&	2.580	\\*
6	&	8.0485	&	4.0	&	0.433	&	0.000	&	0.000	&	0.000	&	4.239	&	2.775	&	0.756	&	1.630	&	7.516	&	2.633	&	2.599	&	2.635	&	2.597	\\*
7	&	8.0915	&	4.0	&	0.491	&	0.000	&	0.000	&	0.000	&	4.222	&	2.822	&	0.756	&	1.660	&	7.533	&	2.781	&	2.620	&	2.783	&	2.618	\\*
8	&	8.1212	&	4.0	&	0.552	&	0.000	&	0.617	&	0.000	&	4.202	&	2.863	&	0.756	&	1.717	&	7.557	&	2.932	&	2.636	&	2.934	&	2.636	\\*
9	&	8.1361	&	4.0	&	0.580	&	0.000	&	0.618	&	0.000	&	4.194	&	2.891	&	0.756	&	1.810	&	7.589	&	3.024	&	2.649	&	3.026	&	2.648	\\*
10	&	8.1420	&	4.0	&	0.524	&	0.000	&	0.619	&	0.000	&	4.236	&	2.946	&	0.756	&	2.293	&	7.688	&	3.071	&	2.673	&	3.074	&	2.672	\\*
11	&	8.1421	&	4.0	&	0.527	&	0.000	&	0.620	&	0.000	&	4.231	&	2.931	&	0.756	&	2.427	&	7.662	&	3.041	&	2.668	&	3.044	&	2.667	\\*
12	&	8.1422	&	4.0	&	0.538	&	0.000	&	0.620	&	0.000	&	4.223	&	2.921	&	0.756	&	2.519	&	7.648	&	3.034	&	2.670	&	3.038	&	2.669	\\*
13	&	8.1427	&	4.0	&	0.643	&	0.000	&	0.624	&	0.000	&	4.187	&	2.989	&	0.756	&	2.879	&	7.632	&	3.325	&	2.693	&	3.327	&	2.693	\\*
14	&	8.1437	&	4.0	&	0.739	&	0.000	&	0.632	&	0.000	&	4.147	&	3.018	&	0.756	&	3.224	&	7.670	&	3.536	&	2.706	&	3.539	&	2.705	\\*
15	&	8.1445	&	4.0	&	0.841	&	0.000	&	0.638	&	0.000	&	4.099	&	3.031	&	0.756	&	3.488	&	7.739	&	3.739	&	2.717	&	3.742	&	2.716	\\*
16	&	8.1449	&	4.0	&	0.940	&	0.000	&	0.640	&	0.000	&	4.050	&	3.034	&	0.756	&	3.667	&	7.798	&	3.929	&	2.730	&	3.932	&	2.729	\\*
17	&	8.1453	&	4.0	&	1.041	&	0.000	&	0.642	&	0.000	&	3.999	&	3.031	&	0.756	&	3.797	&	7.844	&	4.114	&	2.747	&	4.119	&	2.746	\\*
18	&	8.1455	&	4.0	&	1.144	&	0.000	&	0.643	&	0.000	&	3.945	&	3.022	&	0.756	&	3.894	&	7.880	&	4.292	&	2.767	&	4.301	&	2.766	\\*
19	&	8.1456	&	4.0	&	1.248	&	0.000	&	0.644	&	0.000	&	3.890	&	3.008	&	0.756	&	3.967	&	7.908	&	4.455	&	2.785	&	4.482	&	2.781	\\*
20	&	8.1458	&	4.0	&	1.349	&	0.000	&	0.644	&	0.000	&	3.835	&	2.991	&	0.756	&	4.021	&	7.928	&	4.603	&	2.788	&	4.665	&	2.780	\\*
21	&	8.1458	&	4.0	&	1.449	&	0.000	&	0.644	&	0.000	&	3.780	&	2.970	&	0.756	&	4.062	&	7.944	&	4.752	&	2.772	&	4.871	&	2.759	\\*
22	&	8.1459	&	4.0	&	1.538	&	0.000	&	0.645	&	0.000	&	3.728	&	2.942	&	0.756	&	4.096	&	7.957	&	4.747	&	2.770	&	4.956	&	2.747	\\*
23	&	8.1460	&	4.0	&	1.565	&	0.249	&	0.645	&	0.000	&	3.700	&	2.883	&	0.756	&	4.154	&	7.979	&	2.082	&	3.763	&	2.421	&	3.763	\\*
24	&	8.1462	&	4.0	&	1.643	&	1.318	&	0.645	&	0.000	&	3.686	&	2.982	&	0.756	&	4.211	&	8.001	&	1.173	&	3.609	&	1.339	&	3.448	\\*
25	&	8.1467	&	4.0	&	1.733	&	2.305	&	0.647	&	0.000	&	3.677	&	3.125	&	0.753	&	4.206	&	8.096	&	1.091	&	3.522	&	1.265	&	3.328	\\*
26	&	8.1632	&	4.0	&	1.696	&	0.647	&	0.859	&	0.000	&	3.686	&	3.089	&	0.753	&	3.878	&	8.126	&	1.613	&	3.682	&	1.884	&	3.564	\\*
27	&	8.1680	&	4.0	&	1.698	&	0.139	&	0.902	&	0.000	&	3.693	&	3.124	&	0.753	&	3.839	&	8.134	&	3.688	&	3.871	&	4.503	&	3.273	\\*
28	&	8.1961	&	4.0	&	1.336	&	0.000	&	1.149	&	0.000	&	3.920	&	3.303	&	0.753	&	3.693	&	8.200	&	5.975	&	2.904	&	6.000	&	2.902	\\*
29	&	8.2039	&	4.0	&	1.434	&	0.000	&	1.203	&	0.745	&	3.877	&	3.330	&	0.753	&	3.752	&	8.248	&	6.463	&	2.934	&	6.528	&	2.929	\\*
30	&	8.2056	&	4.0	&	1.537	&	0.000	&	1.214	&	0.761	&	3.827	&	3.334	&	0.753	&	3.805	&	8.271	&	6.854	&	2.945	&	6.992	&	2.937	\\*
31	&	8.2064	&	4.0	&	1.638	&	0.000	&	1.219	&	0.774	&	3.776	&	3.333	&	0.753	&	3.849	&	8.288	&	7.254	&	2.941	&	7.518	&	2.928	\\*
32	&	8.2068	&	4.0	&	1.736	&	0.000	&	1.221	&	0.766	&	3.726	&	3.329	&	0.753	&	3.885	&	8.300	&	7.613	&	2.930	&	8.147	&	2.908	\\*
33	&	8.2074	&	4.0	&	1.835	&	1.050	&	1.225	&	0.757	&	3.673	&	3.315	&	0.753	&	4.183	&	8.385	&	1.582	&	3.588	&	1.906	&	3.435	\\*
34	&	8.2075	&	4.0	&	1.925	&	1.922	&	1.225	&	0.752	&	3.662	&	3.452	&	0.753	&	4.509	&	8.431	&	1.378	&	3.468	&	1.693	&	3.265	\\*
35	&	8.2076	&	4.0	&	1.910	&	1.824	&	1.225	&	0.755	&	3.664	&	3.428	&	0.753	&	4.600	&	8.435	&	1.381	&	3.484	&	1.691	&	3.284	\\*
36	&	8.2076	&	4.0	&	1.937	&	2.055	&	1.224	&	0.757	&	3.661	&	3.471	&	0.753	&	4.758	&	8.457	&	1.357	&	3.453	&	1.673	&	3.243	\\*
37	&	8.2078	&	4.0	&	2.036	&	2.469	&	1.223	&	0.768	&	3.651	&	3.629	&	0.751	&	5.130	&	8.504	&	1.375	&	3.369	&	1.754	&	3.132	\\*
38	&	8.2080	&	4.0	&	2.134	&	2.641	&	1.222	&	0.798	&	3.641	&	3.784	&	0.747	&	5.403	&	8.536	&	1.459	&	3.281	&	1.937	&	3.029	\\*
39	&	8.2081	&	4.0	&	2.235	&	2.718	&	1.221	&	0.833	&	3.630	&	3.943	&	0.744	&	5.740	&	8.572	&	1.611	&	3.189	&	2.251	&	2.925	\\*
40	&	8.2082	&	4.0	&	2.337	&	2.762	&	1.219	&	0.867	&	3.618	&	4.100	&	0.741	&	6.135	&	8.582	&	1.846	&	3.094	&	2.766	&	2.828	\\*
41	&	8.2083	&	4.0	&	2.435	&	2.945	&	1.058	&	0.900	&	3.608	&	4.253	&	0.700	&	6.572	&	8.487	&	2.197	&	2.989	&	3.621	&	2.720	\\*
42	&	8.2084	&	4.0	&	2.532	&	3.039	&	0.961	&	0.926	&	3.596	&	4.401	&	0.676	&	6.895	&	8.338	&	4.041	&	2.893	&	8.702	&	2.589	\\*
43	&	8.2084	&	4.0	&	2.562	&	3.046	&	0.954	&	0.940	&	3.592	&	4.445	&	0.675	&	7.025	&	8.280	&	5.678	&	2.835	&	10.594	&	2.531	\\*
44	&	8.2084	&	4.0	&	2.474	&	3.046	&	0.953	&	0.950	&	3.603	&	4.315	&	0.675	&	7.111	&	8.244	&	2.524	&	2.973	&	5.053	&	2.701	\\*
45	&	8.2084	&	4.0	&	2.533	&	3.044	&	0.956	&	0.954	&	3.596	&	4.402	&	0.675	&	7.209	&	8.192	&	4.054	&	3.006	&	--	&	--	\\*
46	&	8.2085	&	4.0	&	2.630	&	3.037	&	0.963	&	0.962	&	3.583	&	4.543	&	0.675	&	7.300	&	8.134	&	--	&	--	&	--	&	--	\\*
47	&	8.2086	&	4.0	&	2.738	&	3.007	&	0.993	&	0.992	&	3.568	&	4.700	&	0.673	&	7.436	&	8.071	&	--	&	--	&	--	&	--	\\*
48	&	8.2091	&	4.0	&	2.841	&	2.898	&	1.102	&	1.102	&	3.556	&	4.858	&	0.667	&	7.780	&	8.090	&	--	&	--	&	--	&	--	\\*
49	&	8.2098	&	4.0	&	2.929	&	2.633	&	1.367	&	1.367	&	3.550	&	5.010	&	0.665	&	9.278	&	8.439	&	--	&	--	&	--	&	--	\\*
	\enddata
	
\tablecomments{1. This table is available in its entirety in machine-readable form.\\
 2. $M_{\rm KH}$ is the mass at which the stellar inner radius $R_\mathrm{KH}$ is equal to its Roche-lobe radius $R_\mathrm{L}$ for the critical mass ratio $q_\mathrm{ad}$.  $\tilde{M}_{\rm KH}$ has the same meaning as $M_{\rm KH}$ but for isentropic envelope stars. See Figures 4 and 6 in \citet{2010ApJ...717..724G} for detail. The value of $M_\mathrm{1i} - M_{\rm KH}$ or $M_\mathrm{1i} - \tilde{M}_{\rm KH}$ indicate whether a prompt or a delayed dynamical instability occurs.}
\end{deluxetable*}
\newcounter{tbl1}

Table~\ref{4m-tab} summarizes both the initial global and interior physical parameters of our $Z=0.001$ IM stars. Typically, the lower and upper mass limits for the IM star with solar metallicity are around $2.1 M_\sun$ and $8 M_\sun$. We select here from $1.6 M_\sun$ and $10.0 M_\sun$ to cover the metallicity effects and a broader range of mass. 

The key parameters are as follows: k is a mass loss sequence number, $t$ is the age, $M$ is the mass of the initial model, $R$ is the initial radius, $M_{\rm ce}$ is the mass of the convective envelope, $M_{\rm He}$ is the mass of the helium core, where mass fraction of $X_\mathrm{H}$ is 0.15, $M_{\rm C}$ is the mass of the carbon core, where mass fraction of $X_\mathrm{He}$ is 0.25, $T_{\rm e}$ is the effective temperature, $L$ is the stellar luminosity, $X_{\rm s}$ is the surface hydrogen abundance (fraction by mass), $\rho_{\rm c}$ is the central density, $T_{\rm c}$ is the central temperature, $q_{\rm ad}$ is the critical mass ratio for dynamical-timescale mass transfer, $M_{\rm KH}$ is the mass threshold at which $\dot{M} = - M/\tau_{\rm KH}$, $\tilde{q}_{\rm ad}$ is the critical mass ratio for dynamical-timescale mass transfer in the case of an isentropic envelope and $\tilde{M}_{\rm KH}$ is the mass threshold at which $\dot{M} = - M/\tau_{\rm KH}$ in that case. The second line in Table\,\ref{4m-tab} lists accordingly the unit of these physical variables.

\begin{figure}[ht!]
\centering
\includegraphics[scale=0.29]{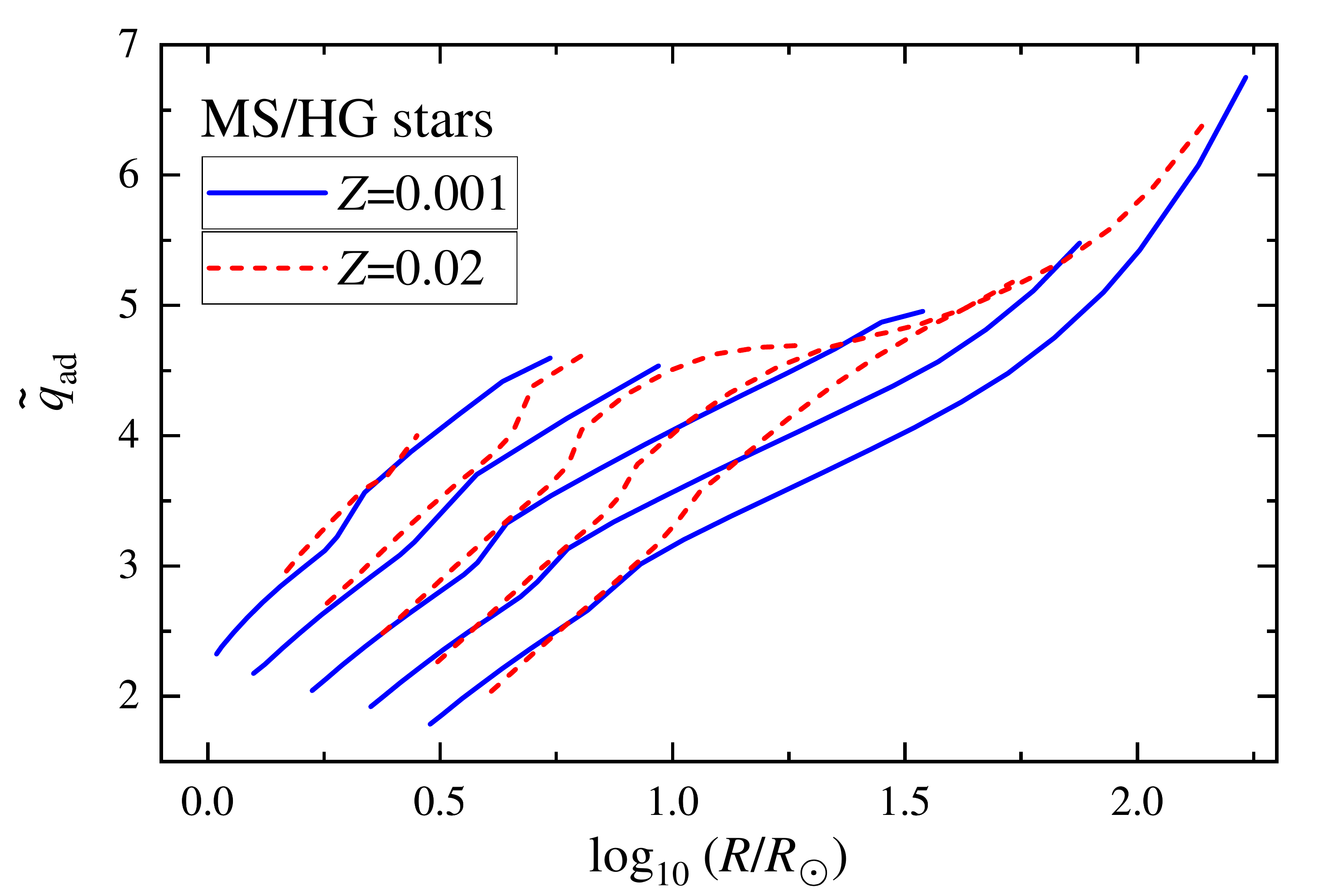}
\caption{Critical mass ratios as a function of the radii of MS/HG donor stars. Blue solid and red dashed lines are metal-poor and solar metallicity stars. From left to right, masses of different lines are $1.6\,M_\sun$, $2.5\,M_\sun$, $4.0\,M_\sun$, $6.3\,M_\sun$ and $10.0\,M_\sun$. For clarity, plots end at the maximum of $\tilde{q}_{\rm ad}$ in the late HG.}
\label{qad-mshg}
\end{figure}
\begin{figure}[ht!]
\centering
\includegraphics[scale=0.29]{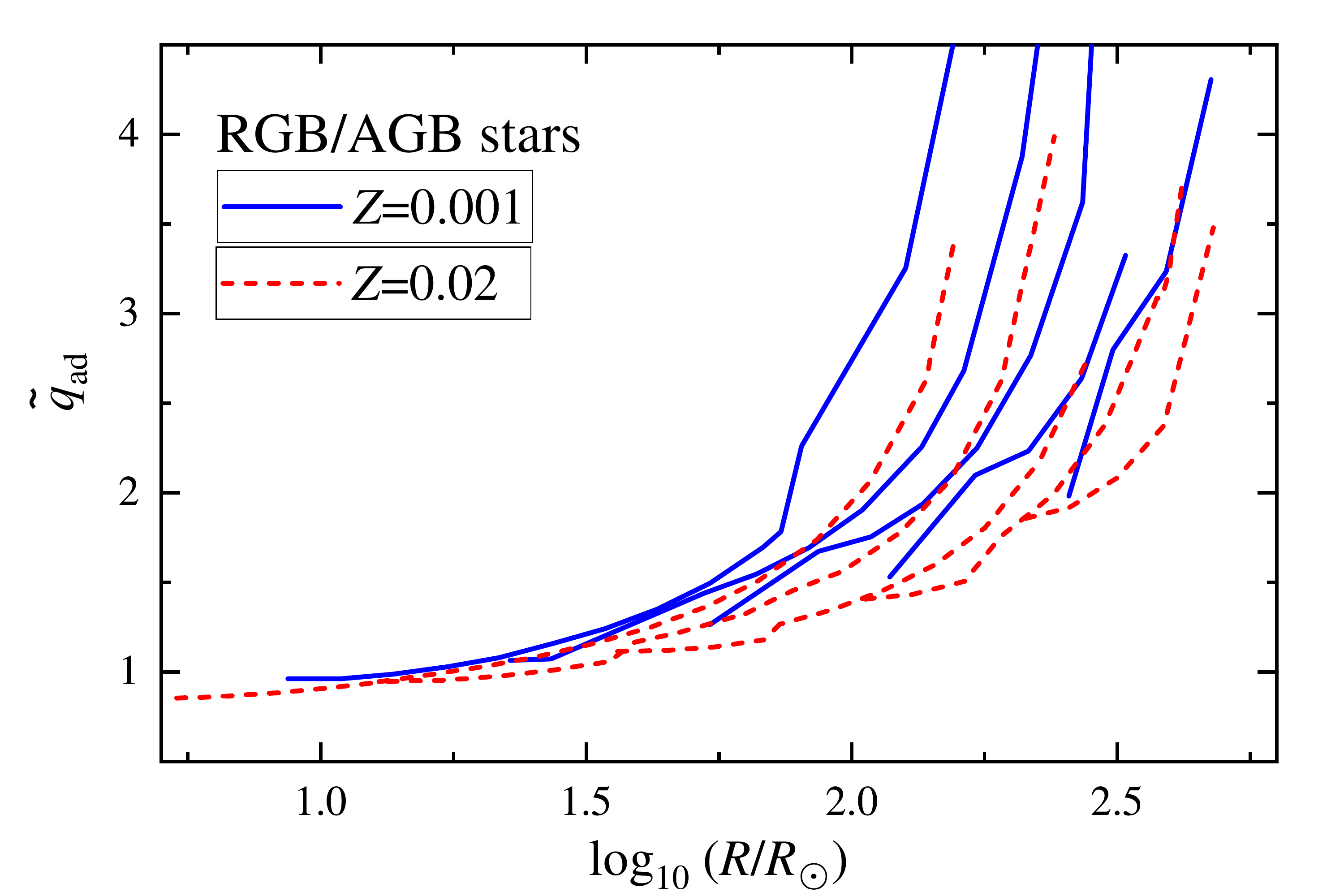}
\caption{Similar with Figure~\ref{qad-mshg} but for RGB/AGB stars. For clarity, plots start from the minimum of $\tilde{q}_{\rm ad}$ on the early RGB.}
\label{qad-gb}
\end{figure}
Critical mass ratios $\tilde{q}_{\rm ad}$ listed in the extended version of Table~\ref{4m-tab} are also partially presented in graphical form in Figures~\ref{qad-mshg} and \ref{qad-gb}. We see that $\tilde{q}_{\rm ad}$ decreases almost linearly with the logarithm of radius at the BRGB region from the late HG to the early RGB. As we found in the last section, $\tilde{q}_{\rm ad}$ is smaller for MS/HG metal-poor stars at the same evolutionary stage. Conversely, $\tilde{q}_{\rm ad}$ is larger for RGB/AGB metal-poor stars. We provide both tabular and graphical forms for our results in this section. We find the fitting formulae for the critical mass ratios as functions of masses and radii in the next section.

\section{Fitting formulae}
\label{sec:fit}

\begin{deluxetable*}{lllllllll}   
	\tabletypesize{\footnotesize}
	\tablewidth{0pt}
	\tablecolumns{9}
	\tablecaption{Fitting formulae and coefficients \label{fit-tab}}
	
	\tablehead{
		\colhead{stage} & \colhead{$Z$} & \colhead{formula} & \colhead{a}
		&\colhead{b} & \colhead{c} & \colhead{$q_\mathrm{min}$} & \colhead{$R_\mathrm{min}$} & \colhead{$R_\mathrm{max}$}
	}
	\startdata
	MS 	&	0.001	&	Eq.~\ref{eq-ms} & 2.88940	&	$-2.46266$	&	2.80378 & --- & --- & --- \\*
	MS 	&	0.02	&	Eq.~\ref{eq-ms}	& 3.09721	&	$-3.05344$	&	3.24722 & --- & --- & --- \\*
	\hline
	MS/HG &	0.001	&	Eq.~\ref{eq-mshg}  & $3.72324 $ 	& $-2.26829$     & $0.19792 $   & Eq.~\ref{eq-mshg-qmin1} & Eq.~\ref{eq-mshg-rmin1} & Eq.~\ref{eq-mshg-rmax1} \\*
	{}   &  {}     &       {}             & $-0.79775\times (M/M_\sun)$   & $+0.56558\times (M/M_\sun)$   & $-0.04750\times (M/M_\sun)$   & {} & {} & {} \\*
	{}   &  {}     &       {}             & $+0.04619\times (M/M_\sun)^2$ & $-0.03254\times (M/M_\sun)^2$ & $+0.00278\times (M/M_\sun)^2$ & {} & {} & {} \\*
	MS/HG &	0.02	&	Eq.~\ref{eq-mshg}  & $0.75208$	& $0.13486$	& $-0.36874$ & Eq.~\ref{eq-mshg-qmin2} & Eq.~\ref{eq-mshg-rmin2} & Eq.~\ref{eq-mshg-rmax2} \\*
	{}   &  {}     &       {}            & $+0.39155\times (M/M_\sun)$ 	& $-0.50319\times (M/M_\sun)$     & $+0.23056\times (M/M_\sun)$ & {} & {} & {} \\*
	{}   &  {}     &       {}             & $-0.03915\times (M/M_\sun)^2$ & $+0.04552\times (M/M_\sun)M^2$ & $-0.03658\times (M/M_\sun)^2$ & {} & {} & {} \\*
	{} & {} & {} & {} & {} &$+0.001732\times (M/M_\sun)^3$ & {} & {} & {} \\*
	\hline
	RGB/AGB &	0.001	&	Eq.~\ref{eq-gb}  & $0.01066$	&	 $-1.18954$E-6	&	$-0.005901$ & Eq.~\ref{eq-gb-qmin1} & Eq.~\ref{eq-gb-rmin1} & --- \\*
	{}   &  {}     &       {}            & $-9.82603$E-4$\times (M/M_\sun)$ 	& $+5.61586$E-6$\times (M/M_\sun)$     & $+0.001507\times (M/M_\sun)$ & {} & {} & {} \\*
	{}   &  {}     &       {}            & $-2.95245$E-4$\times (M/M_\sun)^2$ 	& $+6.02668$E-7$*(M/M_\sun)^2$     & {} & {} & {} & {} \\*
	RGB/AGB &	0.02	&	Eq.~\ref{eq-gb}  & $0.01595$	& $-8.39047$E-5   & $-0.00897$    & Eq.~\ref{eq-gb-qmin2} & Eq.~\ref{eq-gb-rmin2} & --- \\*
	{}   &     {}     &       {}           & $-0.00526\times (M/M_\sun)$ & $+4.21662$E-5$\times (M/M_\sun)$ & $+0.00291\times (M/M_\sun)$   & {} & {} & {} \\*
	{}   &     {}     &       {}           & $+3.64794$E-4$\times (M/M_\sun)^2$ & $-5.14285$E-6$\times (M/M_\sun)^2$ & $-2.19604$E-4$\times (M/M_\sun)^2$   & {} & {} & {} \\*
	{}   &     {}     &       {}           & {} & $+1.77890$E-7$\times (M/M_\sun)^3$ & {} & {} & {} & {} \\*
	\enddata 
	\tablecomments{Min is the abbreviation of minimum, max for maximum and avg for average.}
\end{deluxetable*}

.
The instability criteria for dynamical-timescale mass transfer provide us with onset thresholds for common envelope evolution. This is one of the key physical inputs for binary population synthesis. Interpolation in the tables provide accurate criteria at the cost of calculating speed. Alternatively, we find fitting formulae for both $Z=0.001$ and $Z=0.02$ (Paper III) IM stars with masses from $1.6 M_\sun$ to $10.0 M_\sun$.

\begin{figure}[ht!]
\centering
\includegraphics[scale=0.29]{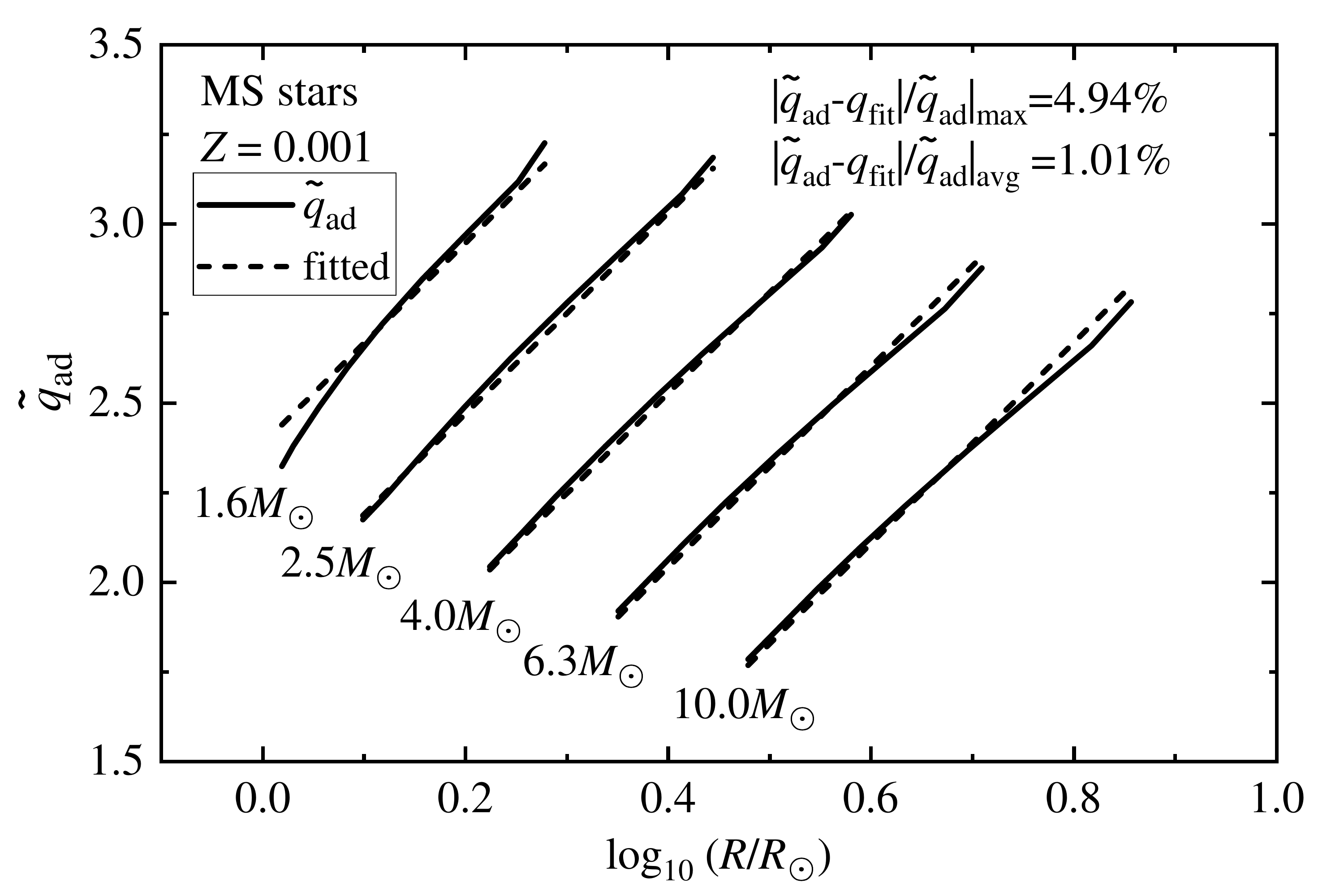}
\caption{Critical mass ratios $\tilde{q}_{\rm ad}$ (solid lines) and the fitted results (dashed lines), functions of the mass and radius for $Z=0.001$ MS stars. These are linearly correlated with the logarithm of mass and radius of the star.}
\label{z001-ms}
\end{figure}

\begin{figure}[ht!]
	\centering
	\includegraphics[scale=0.29]{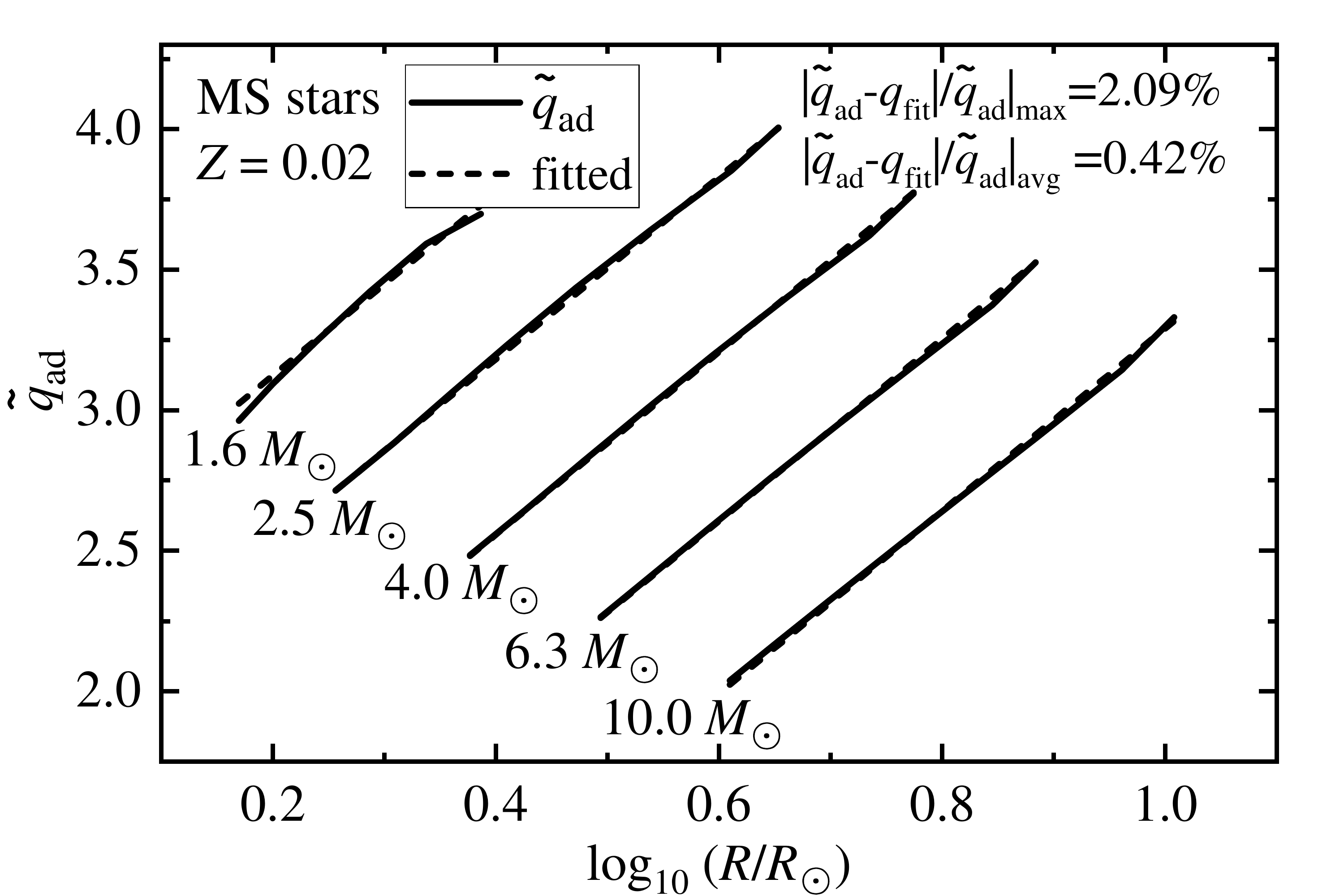}
	\caption{Similar to Figure~\ref{z001-ms} but for $Z=0.02$ MS stars.}
	\label{z02-ms}
\end{figure}

For MS donor stars, there is a linear relationship between the critical mass ratio and the logarithm of mass and radius \citep[see also][]{2013IAUS..290..213G},
\begin{equation}
 \tilde{q}_{\rm ad} = \mathrm{a} + \mathrm{b}\times \log_{10} \left(\frac{M}{M_\sun}\right)+ \mathrm{c}\times \log_{10} \left(\frac{R}{R_\sun}\right).
 \label{eq-ms}
\end{equation} 
The coefficients a, b and c for metal-poor and solar metallicity IM stars are given in Table~\ref{fit-tab} and Figures \ref{z001-ms} and \ref{z02-ms}. We find that the fitting formulas for MS donor stars are simply and accurately fitted. The maximum and average absolute fractional deviation are $4.94\%$ and $1.01\%$ for $Z=0.001$ MS stars. The corresponding values are $2.09\%$ and $0.42\%$ for $Z=0.02$ MS stars.

\begin{figure}[ht!]
	\centering
	\includegraphics[scale=0.29]{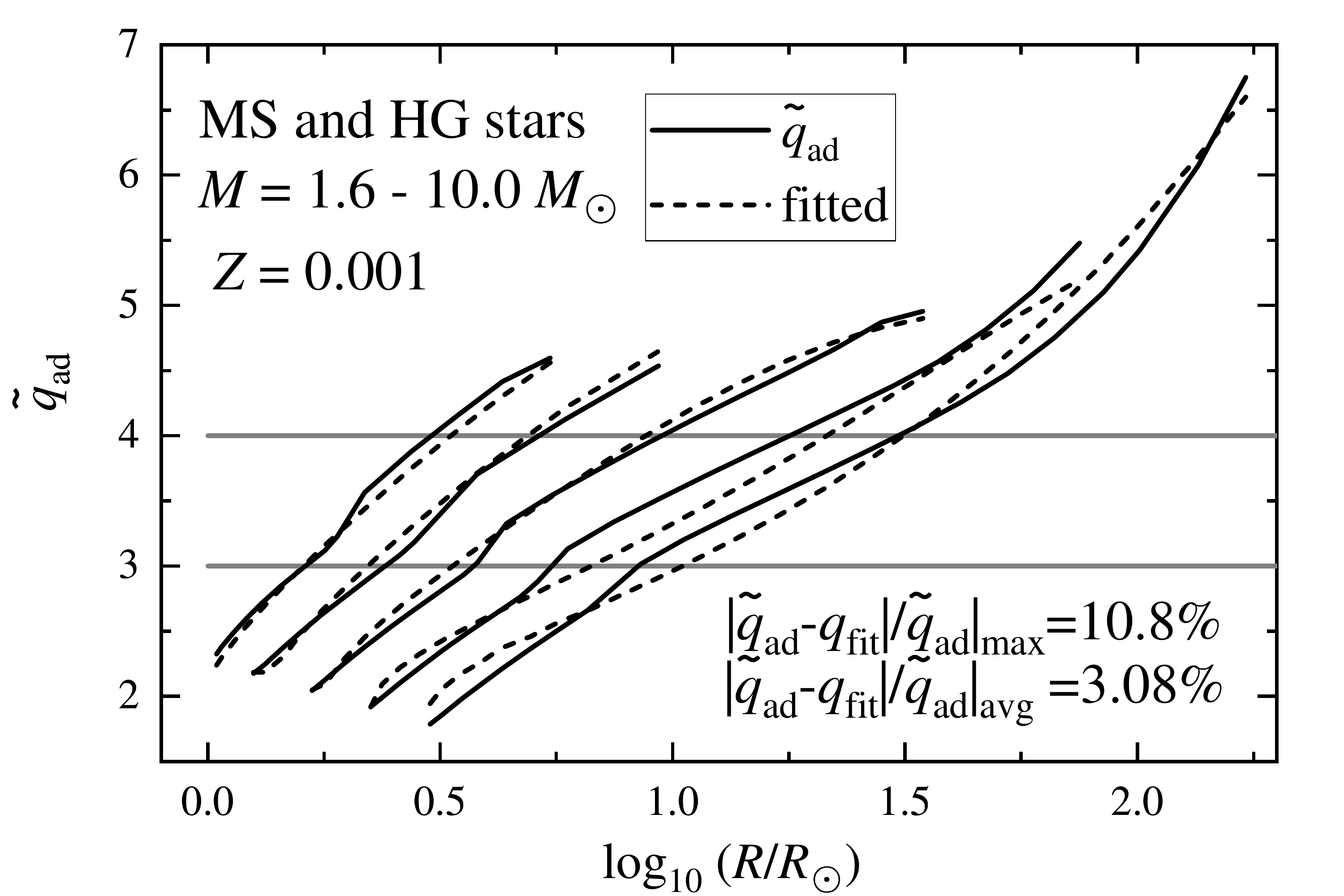}
	\caption{Critical mass ratios (solid lines) and the fitted results (dashed lines) as functions of radius for $Z=0.001$ MS/HG stars. From left to right, masses are $1.6\,M_\sun$, $2.5\,M_\sun$, $4.0\,M_\sun$, $6.3\,M_\sun$ and $10.0\,M_\sun$. Fitting formulae, coefficients and the radius ranges are given in Table~\ref{fit-tab}.}
	\label{z001-mshg}
\end{figure}

\begin{figure}[ht!]
	\centering
	\includegraphics[scale=0.29]{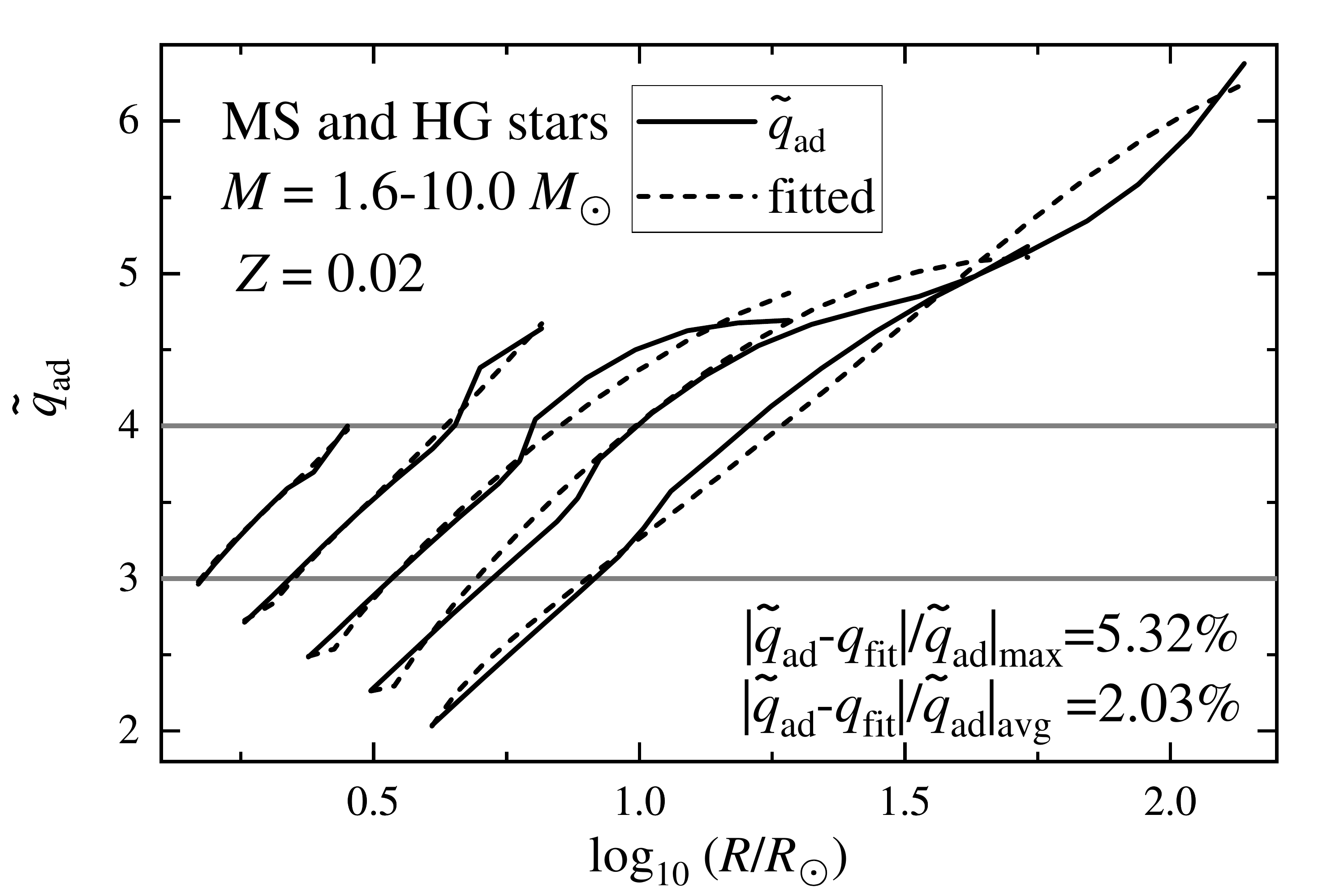}
	\caption{Similar to Figure~\ref{z001-mshg} but for $Z=0.02$ MS/HG stars.}
	\label{z02-mshg}
\end{figure}

The mass fraction of the radiative envelope of the MS/HG donor star increases monotonically from the ZAMS to the late HG. However, gradient of $\tilde{q}_{\rm ad}$ as a function of $\log_{10} R$ differs for HG and MS donor stars (Figure~\ref{qad-mshg}). So, we use the fitting formula for MS/HG stars as follows:
\begin{equation}
	\begin{aligned}
	\tilde{q}_{\rm ad} & = q_\mathrm{min} +\\
	&\frac{\mathrm{a}\times [(R-R_\mathrm{min})/R_\sun]^{1/2}+\mathrm{b}\times [(R-R_\mathrm{min})/R_\sun]^{1/4}}{1+\mathrm{c}\times [(R-R_\mathrm{min})/R_\sun]^{7/8}},
	\end{aligned}
	\label{eq-mshg}
\end{equation} 
where $q_\mathrm{min}$ and $R_\mathrm{min}$ are the critical mass ratio and radius of the ZAMS models. The coefficients a, b and c for MS/HG stars are given in Table~\ref{fit-tab}. Equation (\ref{eq-mshg}) is valid for donor stars with radii $R$ from $R_\mathrm{min}$ to $R_\mathrm{max}$. $R_\mathrm{max}$ is the radius of a donor star at the LHG where the critical mass ratio reaches a maximum. For $Z=0.001$ IM stars, we have
\begin{equation}
q_\mathrm{min} = q_\mathrm{ZAMS} = 2.45122 - 0.66844\times \log_{10} \left(\frac{M}{M_\sun}\right),
	\label{eq-mshg-qmin1}
\end{equation} 
\begin{equation}
\frac{R_\mathrm{min}}{R_\sun} = \frac{R_\mathrm{ZAMS}}{R_\sun} = -0.2922 + \left(\frac{M}{M_\sun}\right)^{0.51},
	\label{eq-mshg-rmin1}
\end{equation} 
and
\begin{equation}
\log_{10} \left(\frac{R_\mathrm{max}}{R_\sun}\right) = 0.29534 + 1.95688\times \log_{10} \left(\frac{M}{M_\sun}\right).
	\label{eq-mshg-rmax1}
\end{equation} 
For $Z=0.02$ IM stars, we have
\begin{equation}
	q_\mathrm{min} = q_\mathrm{ZAMS} = 3.18500 - 1.15243\times  \log_{10} \left(\frac{M}{M_\sun}\right),
	\label{eq-mshg-qmin2}
\end{equation} 
\begin{equation}
	\frac{R_\mathrm{min}}{R_\sun} = \frac{R_\mathrm{ZAMS}}{R_\sun} = -0.09996 + \left(\frac{M}{M_\sun}\right)^{0.60},
	\label{eq-mshg-rmin2}
\end{equation} 
and
\begin{equation}
	\log_{10} \left(\frac{R_\mathrm{max}}{R_\sun}\right) = -0.01035 + 2.15399\times  \log_{10} \left(\frac{M}{M_\sun}\right).
	\label{eq-mshg-rmax2}
\end{equation} 
Figures~\ref{z001-mshg} and \ref{z02-mshg} show the fitted criteria as functions of initial mass and radius of MS/HG stars. The fits are not as good as for the MS but still provide the basic trends. For $Z=0.001$ MS/HG stars the max and average absolute fractional deviation are $10.8\%$ and $3.08\%$. The corresponding values for $Z=0.02$ MS/HG stars are $5.32\%$ and $2.03\%$.

For RGB/AGB IM stars,
\begin{equation}
		\tilde{q}_{\rm ad} =  q_\mathrm{min}
		+\frac{\mathrm{a}\times (R/R_\sun)+\mathrm{b}\times (R/R_\sun)^2}{1+\mathrm{c}\times (R/R_\sun)},
		\label{eq-gb}
\end{equation} 
where $q_\mathrm{min}$ is the minimum critical mass ratio near but slightly after the BRGB.

The coefficients a, b and c for metal-poor and solar metallicity RGB/AGB stars are given in Table~\ref{fit-tab}. Equation (\ref{eq-gb}) is suitable for donor stars with radii $R$ from $R^\mathrm{GB}_\mathrm{min}$ to $R_\mathrm{TRGB}$. For $Z=0.001$ IM stars, we have
\begin{equation}
	q_\mathrm{min} = 0.76856 + 0.12128\times  \left(\frac{M}{M_\sun}\right),
	\label{eq-gb-qmin1}
\end{equation} 
\begin{equation}
	\log_{10} \left(\frac{R^\mathrm{GB}_\mathrm{min}}{R_\sun}\right) = 0.59991  + 1.83362\times \log_{10} \left(\frac{M}{M_\sun}\right).
	\label{eq-gb-rmin1}
\end{equation} 
 For $Z=0.02$ IM stars, we have
\begin{equation}
	q_\mathrm{min} = 0.64877 + 0.12029\times \left(\frac{M}{M_\sun}\right),
	\label{eq-gb-qmin2}
\end{equation} 
\begin{equation}
	\log_{10} \left(\frac{R^\mathrm{GB}_\mathrm{min}}{R_\sun}\right) = 0.31947  + 2.05377\times \log_{10} \left(\frac{M}{M_\sun}\right).
	\label{eq-gb-rmin2}
\end{equation} 

\begin{figure}[ht!]
	\centering
	\includegraphics[scale=0.29]{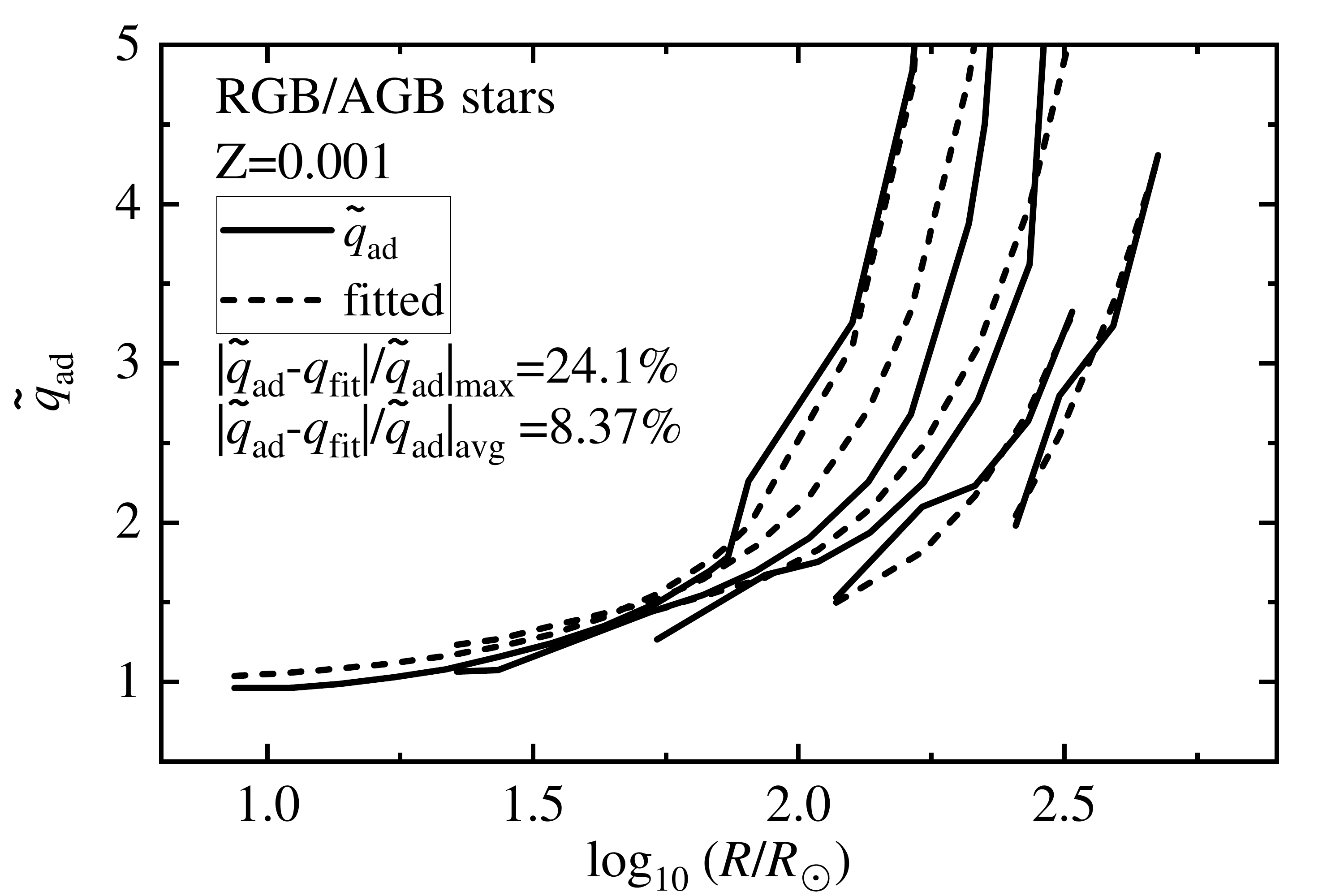}
	\caption{Similar to Figure~\ref{z001-mshg} but for $Z=0.001$ RGB/AGB stars. For late AGB stars, we suggest setting an upper limit around 2-3 for $\tilde{q}_{\rm ad}$.  This is because the critical mass ratio for outer Lagrangian point overflow on a thermal timescale becomes more important \citep[see Figure\,9 by][]{2020ApJS..249....9G}.}
	\label{z001-gb}
\end{figure}
\begin{figure}[ht!]
	\centering
	\includegraphics[scale=0.29]{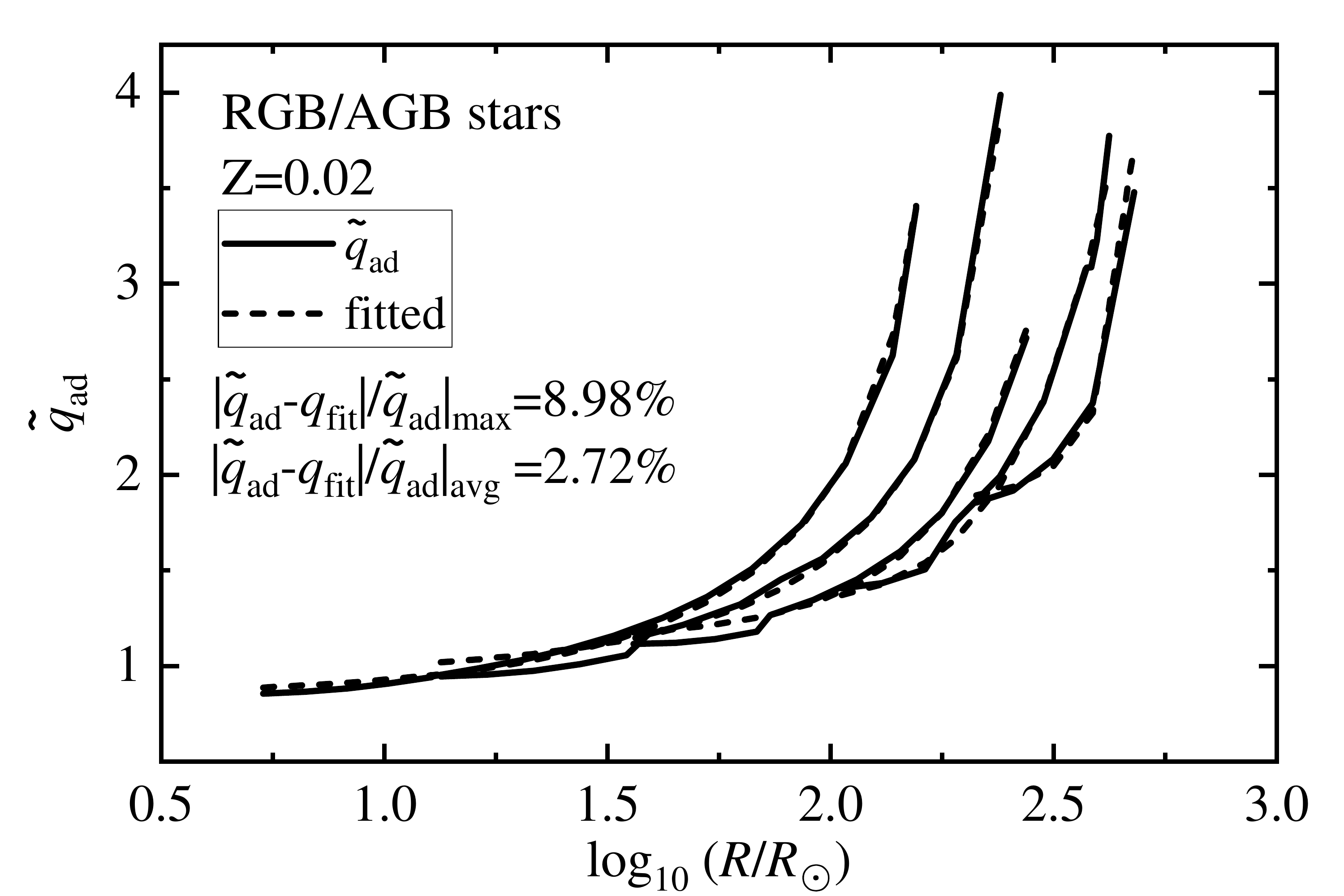}
	\caption{Similar to Figure~\ref{z001-gb} but for $Z=0.02$ RGB/AGB stars.}
	\label{z02-gb}
\end{figure}
Figures~\ref{z001-gb} and \ref{z02-gb} show the fitted criteria as a function of initial mass and radius of RGB/AGB stars. The critical mass ratio $\tilde{q}_{\rm ad}$ increases gradually from less than 1 to larger than 3. This is due to the competition between an increasing convective envelope and the decreasing thermal timescale. We take k=37 and k=38 of $4\,M_\odot$ $Z=0.001$ AGB donor stars as two examples. The convective envelope mass increases from 2.47 to 2.64\,$M_\sun$. We expect the critical mass ratio decreases as the growth of the convective envelope by \citep{1987ApJ...318..794H}. However, the Kelvin–Helmholtz timescale decreases from 1082 to 604 years. Consequently, $\tilde{M}_\mathrm{KH}$ decreases from 3.13 to 3.03\,$M_\sun$ and $\tilde{R}_\mathrm{KH}/\tilde{R}_\mathrm{i}$ decreases from 0.76 to 0.70 at the tangent point (similar with the right panel of Figure\,\ref{4m-z001-q}). So the critical mass ratio $\tilde{q}_\mathrm{ad}$ increases instead from 1.75 to 1.94. The fitting formula's maximum and average absolute fractional deviation are $8.98\%$ and $2.72\%$ for $Z=0.02$ RGB/AGB stars. The accuracy of the fitting formula for $Z=0.001$ RGB/AGB stars is not as good as for solar metallicity stars. The maximum deviation is $24.1\%$ for $2.5\,M_\sun$ stars. But the average deviation of all RGB/AGB stars is acceptable ($8.37\%$).

It is important that $\tilde{q}_{\rm ad}$ drops dramatically around the base of RGB, from the very late HG to the very early RGB (Figure~\ref{4m-zq}). This change is caused by the switch from a radiatively-dominated to a convectively-dominated envelope of the donor star. So, from $R_\mathrm{max}$ of HG to $R^\mathrm{GB}_\mathrm{min}$, $\tilde{q}_{\rm ad}$ can be linearly interpolated by the logarithm of the radius.

\section{Discussions}
\label{sec:imxbs}

It is generally believed that bright Galactic X-ray sources are powered by accreting neutron stars or black holes in binary systems\,\citep[e.g.][]{2006csxs.book..623T}. Among X-ray binary systems, over $90\%$ are high-mass X-ray binaries (HMXBs, donor mass $M_\mathrm{d} \ge 10 M_\sun$) undergoing wind or atmosphere Roche-lobe overflow (RLOF) and low-mass X-ray binaries (LMXBs, donor mass $M_\mathrm{d} \le 1 M_\sun$) suffering Roche-lobe overflow\,\citep{2006csxs.book..623T}. The donor in an IMXB transfers mass to the compact accretor in a thermal/subthermal timescale, and the mass transfer is dynamically stable but non-conserved \citep[e.g.][etc]{2000ApJ...529..946P,2000ApJ...530L..93T,2002ApJ...565.1107P,2012ApJ...756...85S}. Since IMXBs are binary systems that avoid the common envelope process, we use them to compare their mass ratios with critical values for dynamical timescale mass transfer.

We explore the catalogue of LMXBs\,\citep{2003AA...404..301R,2007AA...469..807L}, HMXBs\,\citep{2006AA...455.1165L} and paper about ULXs,\,\citep{2020AA...642A.174M}. A cross-check is made with more extensive catalogs BlackCAT\,\citep{2016A&A...587A..61C}, WATCHDOG\,\citep{2016ApJS..222...15T} and ULXs\,\citep{2022MNRAS.509.1587W}. We pick out 17 IMXBs and candidates with known orbital periods and mass ratios (Table\,\ref{imxbs}). In the following subsections, we first check our theoretical prediction of the critical mass ratios and the observed mass ratios of IMXBs as a function of orbital periods. We then predict the upper and lower X-ray luminosities $L_\mathrm{X}$ of IMXBs and make a comparison with observed IMXBs.

\begin{deluxetable*}{rrrlrrrrrrrrrrr}   
	\tabletypesize{\footnotesize}
	\tablewidth{0pt}
	\tablecolumns{15}
	\rotate
	\tablecaption{Observed IMXBs and candidate IMXBs \label{imxbs}}
	
	\tablehead{
		\colhead{Name1} & \colhead{Name2} & \colhead{$\frac{P_\mathrm{orb}}{\rm d}$} & \colhead{Type$M_\mathrm{X}$}
		&\colhead{SpType$M_\mathrm{d}$} & \colhead{$\log_{10} \frac{L_\mathrm{X}}{\rm erg \,s^{-1}}$} & \colhead{$\sigma_{\log_{10}\frac{L_\mathrm{X}}{\rm erg \,s^{-1}}}$} & \colhead{Ecce} & \colhead{$q=\frac{M_\mathrm{d}}{M_\mathrm{X}}$} & \colhead{$\sigma_q$} & \colhead{$\frac{M_\mathrm{X}}{M_\sun}$}& \colhead{$\sigma_{\frac{M_\mathrm{X}}{M_\sun}}$} & \colhead{$\frac{M_\mathrm{d}}{M_\sun}$} & \colhead{$\sigma_{\frac{M_\mathrm{d}}{M_\sun}}$}& \colhead{Ref}
	}
	\startdata
3A 1909+048	&	SS 433	&	13.100	&	BH	&	A4I-A8I	& $>37.30$	&	---	&	---	&	$\le 1.67$	&	-0.92	&	$\ge$ 5-9	&	---	&	10.40	&	±2.10	&	1-3	\\*
3A 1909+048	&	SS 433	&	13.080	&	BH	&	A7I	&	---	&	---	&	---	&	2.86	&	$^{+1.60}_{-1.10}$	&	4.30	&	±0.80	&	12.30	&	±3.30	&	4	\\*
SAX J1819.3-2525	&	V4641 Sgr	&	2.817	&	BH	&	B9/3	&	39.23	&	---	&	---	&	0.45	&	±0.04	&	6.40	&	±0.60	&	2.90	&	±0.40	&	5	\\*
SAX J1819.3-2525	&	V4641 Sgr	&	2.817	&	BH	&	B9/3	&	39.46 	&	2.30E-01	&	---	&	0.67	&	±0.04	&	9.61	&	$^{+2.08}_{-0.88}$	&	6.53	&	$^{+1.60}_{-1.03}$	&	6	\\*
V1033 Sco	&	GRO J1654-40	&	2.621	&	BH	&	F6/4	&	---	&	---	&	---	&	0.26	&	±0.04	&	5.40	&	±0.30	&	1.45	&	±0.35	&	7	\\*
V1033 Sco	&	GRO J1654-40	&	2.621	&	BH	&	F6/4	&	---	&	---	&	---	&	0.42	&	±0.03	&	6.59	&	±0.45	&	2.76	&	±0.33	&	8	\\*
BW Cir	&	GS 1354-6429	&	2.545	&	BH	&	G0-5/3	&	38.41	&	---	&	---	&	0.12	&	±0.02	&	$\ge 7.83$	& ±0.50	&	$\ge 1.02$	&	±0.17	&	9-10	\\*
IL Lup	&	4U 1543-47	&	1.116	&	BH	&	A2/5	&	39.56 	&	---	&	---	&	0.29	&	$^{+0.21}_{-0.14}$	&	9.40	&	±2.00	&	2.70	&	±1.00	&	11-13\\*
GRO J1716-24	&	V2293 Oph	&	0.613	&	BH	&	---	&	---	&	---	&	---	&	0.33	&	---	&	$\ge 4.90$	&	---	&	1.60	&	---	&	14	\\*
\hline
2S 1417-624	&	---	&	42.120	&	NS	&	B1Ve	&	$>37.34$	&	---	& 0.446	&	$\ge 4.20$	&	---	&	1.40	&	---	&	$\ge 5.90$	&	---	&	15-16	\\*
KS 1947+300	&	GRO J1948+32	&	40.415	&	NS	&	B0Ve	&	38.04 	&	---	&	0.033	&	$\ge 3.57$	&	+3.57?	&	1.40	&	---	&	$\ge 5.00?$	&	+5.00?	&	17	\\*
AX J0049-729	&	RX J0049.1-7250	&	33.380	&	NS	&	B3Ve	&	37.54 	&	---	&	0.400	&	5.36	&	±1.07	&	1.40	&	---	&	7.50	&	±1.50	&	18	\\*
4U 1901+03	&	---	&	22.580	&	NS	&	---	&	38.04 	&	---	&	0.036	&	$>3.21-4.29$	&	---	&	1.40	&	---	& $\ge 4.50$	&	+1.50	&	19	\\*
3A 1909+048	&	SS 433	&	13.100	&	NS?	&	pec	&	---	&	---	&	---	&	4.00	&	±1.18	&	0.80	&	±0.10	&	3.20	&	±0.40	&	20	\\*
SAX J2103.5+4545	&	---	&	12.680	&	NS	&	B0Ve	&	35.90 	&	---	&	0.400	&	5.00	&	---	&	1.40	&	---	&	7.00	&	---	&	21	\\*
2A 1655+353	&	Her X-1	&	1.700	&	NS	&	A9-B	&	37.30 	&	---	&	---	&	2.03	&	±0.42	&	0.98	&	±0.12	&	1.99	&	±0.14	&	22	\\*
RX J0050.7-7316	&	AX J0051-733	&	1.416	&	NS	&	---	&	36.30 	&	---	&	---	&	6.66	&	±1.67	&	1.40	&	---	&	8.70	&	---	&	23	\\*
RX J0050.7-7316	&	AX J0051-733	&	1.416	&	NS	&	---	&	36.30 	&	---	&	---	&	2.94	&	---	&	1.40	&	---	&	4.12	&	---	&	23	\\*
\hline
1WGA J0648.0-4419	&	HD 49798	&	1.550	&	NS	&	sdO6	&	32.00	&	---	&	---	&	1.17	&	±0.09	&	1.28	&	±0.05	&	1.50	&	±0.05	&	24-26	\\*
\hline
NGC 5907 ULX1	&	---	&	5.300	&	NS-ULXs	&	---	&	40.88 	&	4.62E-01	&	---	&	2.86	&	±1.43	&	1.40	&	---	&	4.00	&	±2.00	&	27-28	\\*
M82 X-2	&	---	&	2.533	&	NS-ULXs	&	---	&	39.82 	&	7.00E-03	&	$\leq 0.003$	&	$\geq 3.71$	&	+2	&	1.40	&	---	&	$\geq 5.20$	&	+2.80	&	29-30	\\*
M82 X-2	&	---	&	2.520	&	NS-ULXs	&	---	&	39.82 	&	7.00E-03	&	---	&	3.93	&	±1.79	&	1.40	&	---	&	5.50	&	±2.50	&	31	\\*
M51 ULX-7	&	---	&	1.997	&	NS-ULXs	&	---	&	39.85 	&	7.00E-02	&	$\leq 0.220$	&	$\geq 5.71$	&	---	&	1.40	&	---	&	$\geq 8.00$	&	---	& 32 \\*
	\enddata 
	\tablecomments{Ref\\
	1.\,\citet{2019MNRAS.485.2638C}; 2.\,\citet{2021MNRAS.506.1045M};
	3.\,\citet{2019AA...624A.127W}; 4.\,\citet{2008ApJ...676L..37H}; 5.\,\citet{2014ApJ...784....2M}; 6.\,\citet{2001ApJ...555..489O}; \\
	7.\,\citet{2002MNRAS.331..351B}; 8.\,\citet{2003MNRAS.339.1031S}; 
	9.\,\citet{2009ApJS..181..238C}; 10.\,\citet{2004ApJ...613L.133C}; 
    11.\,\citet{1998ApJ...499..375O}; 12.\,\citet{2003IAUS..212..365O}; \\
    13.\,\citet{2004ApJ...610..378P}; 14.\,\citet{1996AA...314..123M};  
    15.\,\citet{1996AAS..120C.209F}; 16.\,\citet{2004MNRAS.349..173I}
    17.\,\citet{2004ApJ...613.1164G}; 18.\,\citet{2011MNRAS.416.1556T}; \\ 19.\,\citet{2005ApJ...635.1217G}; 20.\,\citet{1991Natur.353..329D}; 21.\,\citet{2000ApJ...544L.129B}; 22.\,\citet{1989PASJ...41....1N}; 23.\,\citet{2000MNRAS.311..169C}; 24.\,\citet{2017ApJ...847...78B}; \\ 25.\,\citet{2009Sci...325.1222M}; 26.\,\citet{2021MNRAS.504..920M};
    27.\,\citet{2020AA...642A.174M}; 28.\,\citet{2017Sci...355..817I}; 29.\,\citet{2014Natur.514..202B}; 30.\,\citet{2022ApJ...937..125B}; \\ 31.\,\citet{2015ApJ...802L...5F}; 32.\,\citet{2020ApJ...895...60R}.}
    \tablecomments{Objects are selected from the catalogs of LMXBs \citep{2003AA...404..301R,2007AA...469..807L}, HMXBs \citep{2006AA...455.1165L} and the paper about ultra-luminous X-ray binaries \citep{2020AA...642A.174M}.}
\end{deluxetable*}

\subsection{Mass ratios of IMXBs}
\label{q-imxbs}

\begin{figure}[ht!]
	\centering
	\includegraphics[scale=0.32]{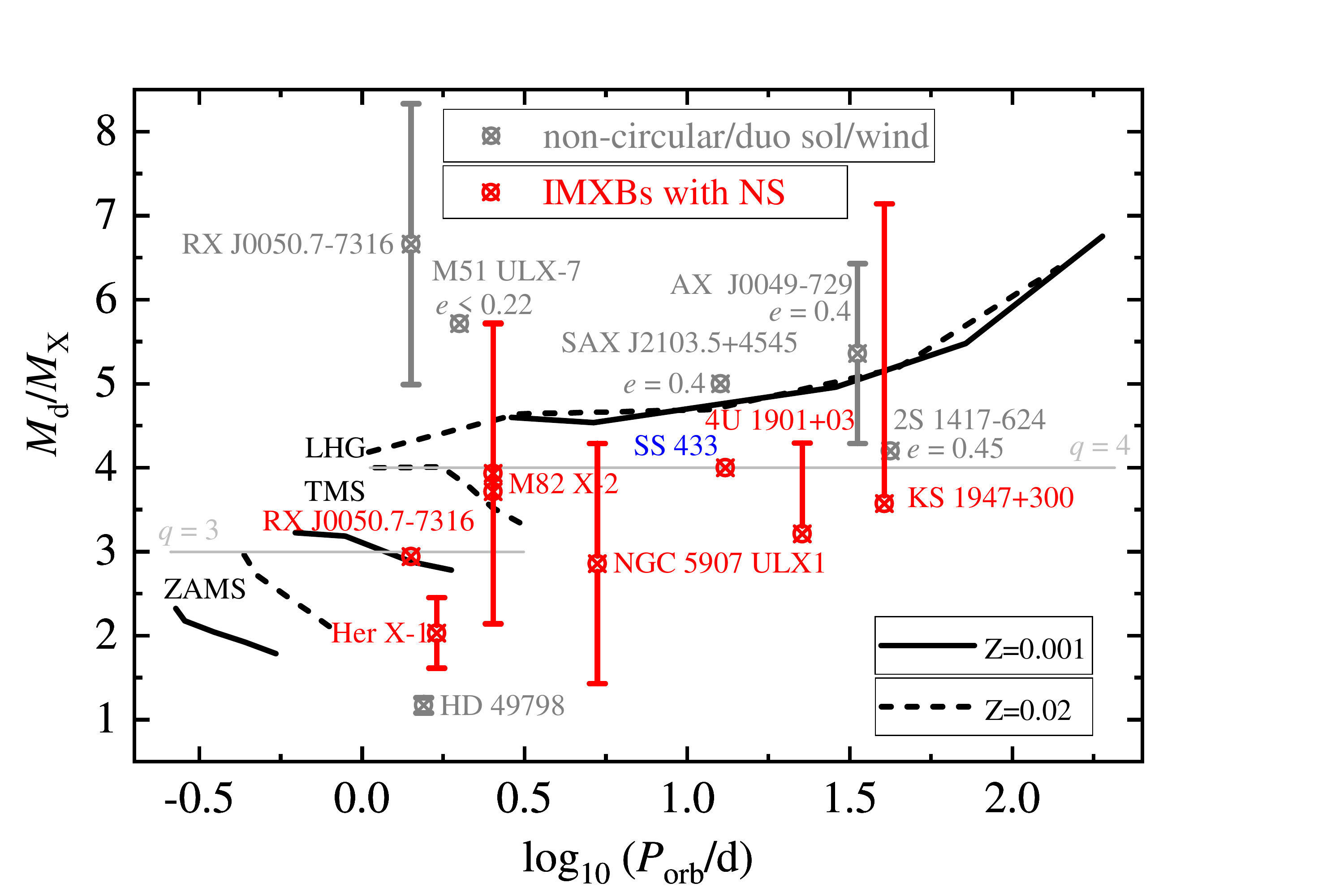}
	\caption{Mass ratio as a function of orbital period for intermediate-mass X-ray binaries (IMXBs) containing a neutron star accretor. Black solid ($Z=0.001$) and dashed ($Z=0.02$) lines show the mass ratio $M_\mathrm{d}/M_\mathrm{X}=\tilde{q}_\mathrm{ad}(M_\mathrm{d},R)$ where the radius $R$ of the donor euqals its Roche-lobe radius $R_\mathrm{L}$ for the given orbital period $P_\mathrm{orb}$. From bottom to up, black solid and dashed lines are for donor stars on the ZAMS, TMS and LHG. Red symbols are IMXBs with small eccentricities. These are located right below our predictions (black lines). Grey symbols are eccentric IMXBs, which cannot be constrained directly from our 1D model. The best-fit mass ratio of RX J0050.7-7316 fits better but others suggest its orbital period might be around 150 d.}
	\label{imxb-ns}
\end{figure}

\begin{figure}[ht!]
	\centering
	\includegraphics[scale=0.32]{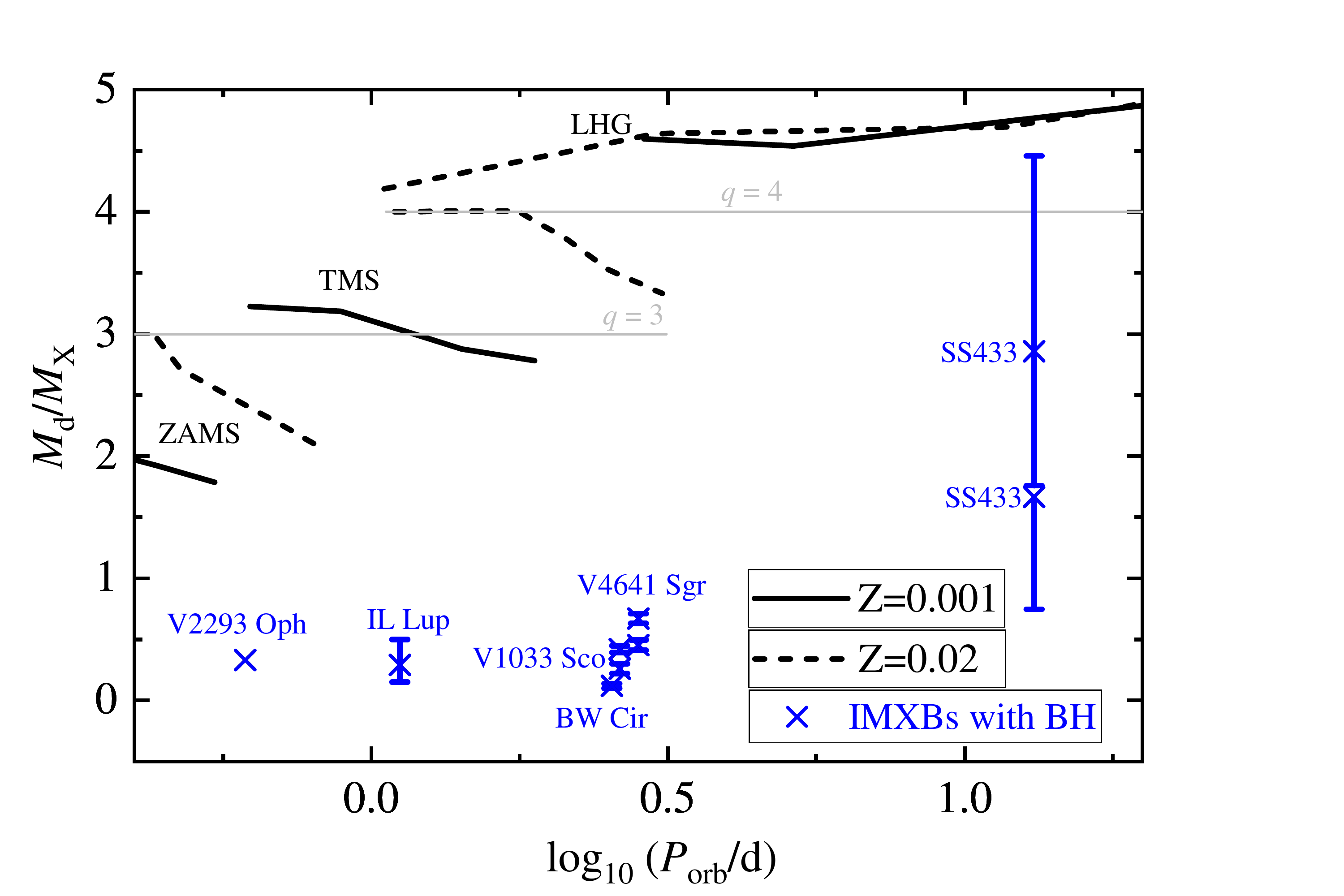}
	\caption{Similar to Figure~\ref{imxb-ns} but for IMXBs containing a black hole (blue cross).}
	\label{imxb-bh}
\end{figure}

We assume all IMXBs are undergoing RLOF. This assumption should be valid for most objects, although some might only fill around $90\%$ of the Roche lobes. We plot the critical mass ratios $\tilde{q}_\mathrm{ad}$ of IM stars on the ZAMS, TMS and LHG ($\tilde{q}^{\rm max}_\mathrm{ad}$) as a function of orbital period $P_\mathrm{orb}$ as solid ($Z=0.001$) and dashed ($Z=0.02$) lines in Figures\,\ref{imxb-ns} and \ref{imxb-bh}. If $q \ge \tilde{q}_\mathrm{ad}$ when the donor first fills its Roche lobe delayed dynamical-timescale mass transfer and common envelope evolution would have altered the system. Thus IMXBs should all have $q < \tilde{q}_\mathrm{ad}$ now to have survived. Mass ratios $q \equiv M_\mathrm{d}/M_\mathrm{X}$ of observed IMXBs are nicely located under the critical mass ratio limit, except for those with an eccentric orbit. We find our prediction is consistent with both the shorter and longer orbital period IMXBs. Compared with the constant critical mass ratios $q=4$ for HG stars, our parameters ($P_\mathrm{orb}$ and $q$) space to form longer period ($P_\mathrm{orb} > 1$d) IMXBs is slightly larger. On the contrary, our parameters space to form shorter period ($P_\mathrm{orb} < 3$d) IMXBs with MS donors is marginally smaller.


We need to keep in mind that mass ratio determination is less accurate than that of the orbital period. Absolute masses of the donor star and the compact accretor are less accurate than the mass ratios. Hence, multiple mass ratio estimates exist for the same object. The neutron star mass might be too low for SS433\,\citep{1991Natur.353..329D} and Her X-1\,\citep{1989PASJ...41....1N}, but the mass ratio should be meaningful. Accretor in SS\,433 is not definitely known, but broadly accepted to be a BH. SS\,433 is one of the youngest X-ray binaries \citep[][and references therein]{2020RAA....20..162L}. This tends to explain why the mass ratio of SS\,433 is larger than most of the observed IMXBs with a BH companion. Only lower limits for the mass ratio of 2S~1417-624 \citep{1996AAS..120C.209F} and M82~X-2 \citep{2014Natur.514..202B} are available. In addition to the minimum value, the donor mass of KS~1947+300 \citep{2004ApJ...613.1164G} could be at least up to $10\,M_\sun$, implying an inclination of $38^{\circ}$. The most probable donor mass (for an inclination of $60^{\circ}$) of 4U\,1901+03 \citep{2005ApJ...635.1217G} is $6\,M_\sun$. \citet{2005ApJ...635.1217G} mention the neutron star in 4U\,1901+03 probably accretes from the wind of an MS OB star. But the X-ray luminosity $L_{\rm X} = 1.1\times 10^{38}$ ${\rm erg\,s^{-1}}$ is high enough, and the donor star can overfill its Roche lobe at the HG. So, we keep this source in Figure\,\ref{imxb-ns}. The X-ray luminosity of HD 49798 IMXB is quite low of $\sim 1.0\times 10^{32}$ ${\rm erg\,s^{-1}}$. The hot subdwarf donor of HD 49798 might be undergoing wind mass transfer at a rate of $ 2.1\times 10^{-9} \,M_\sun \,\rm yr^{-1}$ \citep{2021MNRAS.504..920M}. So we mark this object as gray in Figure\,\ref{imxb-ns}. The mean mass for the donor of M82\,X-2 \citep{2022ApJ...937..125B} is $8\,M_\sun$. RX~J0050.7-7316 (AX~J0051-733; \citealt{2000MNRAS.311..169C}) seems to be the most debatable object. The best-fit mass ratio $q=2.94$ by \citet{2000MNRAS.311..169C} locates well below our predicted upper limit. But the spectrum of the donor star also supports a larger mass ratio $q=6.66$ \citep[][and reference therein]{2000MNRAS.311..169C}. We suspect that the mass of a stripped star could be overestimated based on its spectrum if it is not in thermal equilibrium. \citet{2005MNRAS.362..952C} and \citet{2005AJ....130.2220S} argue that the observed period might actually be the non-radial pulsation and the X-ray data suggest a much longer orbital period of 108 d \citep{2005ApJS..161...96L} or 185 d \citep{1999PASJ...51L..15I}. 

We have not considered eccentric orbits when we calculated critical mass ratios, which are derived on the assumption that $e=0$. So our results are not valid for eccentric IMXBs, such as 2S~1417-624, AX~J0049-729, SAX~J2103.5+4545, and M51~ULX-7. Our critical mass ratios are given for IMXBs that initially formed and triggered RLOF. However, mass ratios of observed IMXBs decrease gradually during the thermal timescale mass transfer process. For IMXBs with a black hole, most of these objects' mass ratios are reversed (less than one) except SS 433. This is consist with our expectation that lower mass transfer rate after the mass ratio reverse lasts a long time to be observed.

\subsection{Max X-ray luminosities of IMXBs}

The accretion luminosity of accreting black holes may be written as \citep{2002apa..book.....F},
\begin{equation}
\begin{aligned}
L_\mathrm{acc} & =2 \eta_\mathrm{acc} G M_\mathrm{X} \dot{M}_\mathrm{acc}/R_\ast \\
               & =\eta_\mathrm{acc} \dot{M}_\mathrm{acc} c^2,
\end{aligned}
\label{Lacc}
\end{equation}
where the dimensionless parameter $\eta_\mathrm{acc}$ measures how efficiently the rest mass energy, $c^2$ per unit mass, of the accreted material is converted into radiation, $R_\ast = 2G M_\mathrm{X} / c^2$ defines the black hole radius. The dimensionless efficiency parameter is generally taken to be $\eta_\mathrm{acc}=0.1$, but it could be up to $\eta_\mathrm{acc}=0.2$ or 0.4 for a $1.4M_\sun$ neutron star or a maximally rotating BH. The Eddington limit to accretion luminosity is \citep{2002apa..book.....F}
\begin{equation}
\begin{aligned}
L_\mathrm{Edd} & = 4\pi G M_\mathrm{X} m_\mathrm{p} c/\sigma_\mathrm{T} \\
		       & \approx 1.3\times 10^{38} (M_\mathrm{X}/M_\sun)\,\mathrm{erg\,s^{-1}},
\end{aligned}
\label{Ledd}
\end{equation}
where $m_\mathrm{p}$ is the proton mass and $\sigma_\mathrm{T} = 6.7 \times 10^{-25} \mathrm{cm^2}$ is the Thomson cross section for fully ionized hydrogen. The corresponding Eddington limit to mass accretion rate \citep{2020AA...642A.174M} is
\begin{equation}
\dot{M}_\mathrm{Edd} \approx 1.5 \times 10^{-8} (M_\mathrm{X}/1.3M_\sun) M_\sun \mathrm{yr}^{-1}
\label{Medd}
\end{equation}

\begin{figure}[ht!]
	\centering
	\includegraphics[scale=0.29]{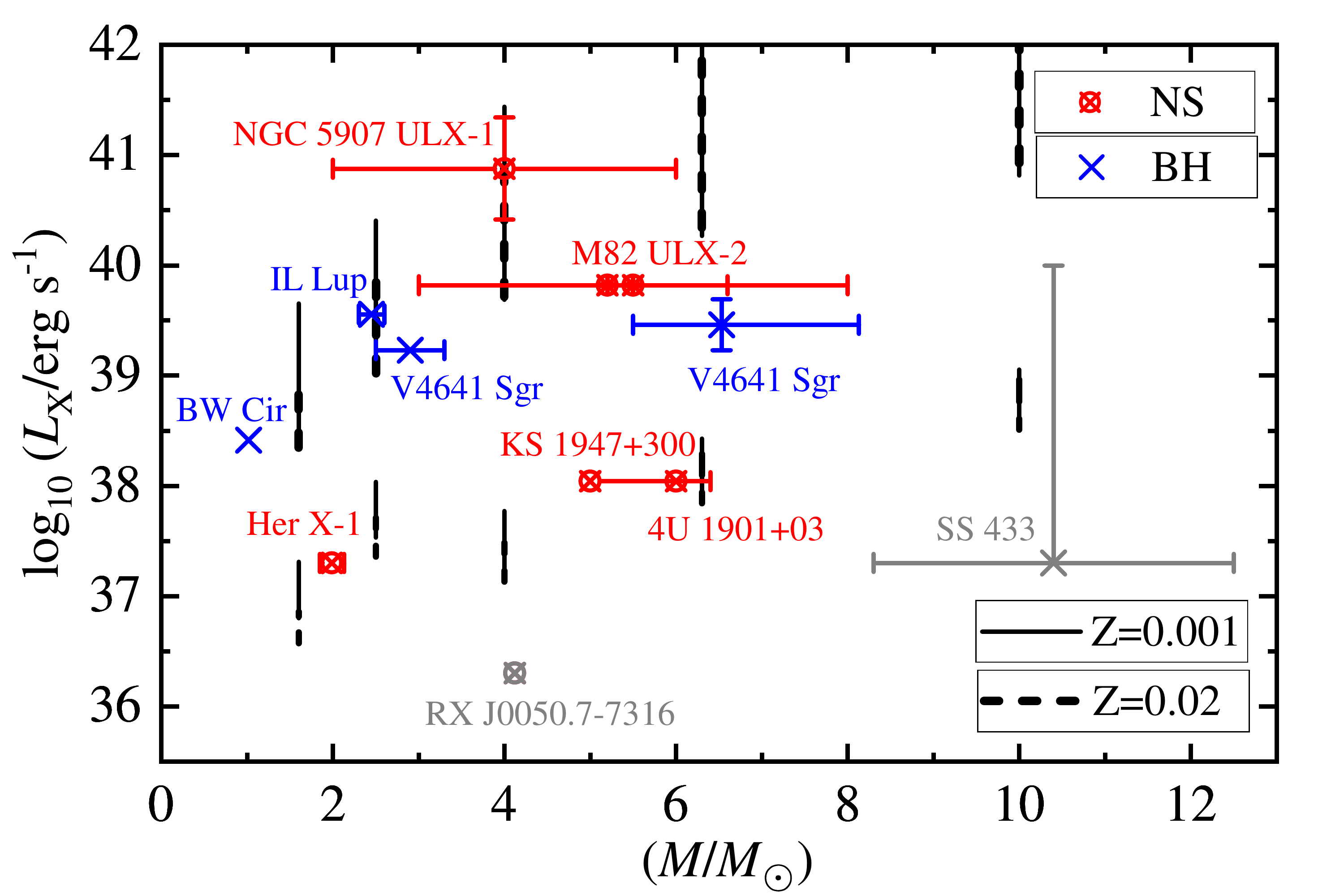}
	\caption{X-ray peak luminosity of IMXBs as a function of donor mass. Red and blue symbols indicate IMXBs with an NS or BH accretor. Black solid ($Z=0.001$) and dashed ($Z=0.02$) vertical lines are theoretical predictions. The masses of these black vertical lines are 1.6, 2.5, 4.0, 6.3, and 10.0\,$M_\sun$ from left to right. According to the mass transfer rate of the donor, there are two track zones. The upper tracks of $L_\mathrm{X}$ follow $0.02 \dot{M}^\mathrm{d}_\mathrm{KH} c^2$ (non-conserved thermal timescale mass transfer and super-Eddington accretion), while the lower tracks obey $0.1 \dot{M}^\mathrm{d}_\mathrm{nuc} c^2$ (conserved nuclear timescale mass transfer) or $0.01 \dot{M}^\mathrm{d}_\mathrm{nuc} c^2$ (non-conserved nuclear timescale mass transfer and super-Eddington accretion). RX J0050.7-7316 is a problematic object but see discussions in Subsection\,\ref{q-imxbs}.}
	\label{lx-imxb}
\end{figure}

To explain the X-ray spectrum of IMXBs, low/hard or the high/soft states \citep{2006ARA&A..44...49R}, requires detailed accretion physics with disk formation, angular momentum transfer, magnetic fields, energy dissipation (collisions of gas elements, shocks, viscous dissipation, etc. \citealt{2002apa..book.....F}). However, the overall accretion energy of the compact accretor is generated by the material transferred from the donor star. Our predictions for the extremes of the X-ray luminosity are consistent with the observed systems (Figure\,\ref{lx-imxb}).

Thermal timescale mass transfer rate (paper I) of the donor star can be written as
\begin{equation}
	\dot{M}^\mathrm{d}_\mathrm{KH} = M_\mathrm{d}/\tau_\mathrm{KH} = R_\mathrm{d} L_\mathrm{d}/ G M_\mathrm{d},
	\label{MKH}
\end{equation}
where $\tau_\mathrm{KH}$ is the Kelvin-Helmholtz timescale and the nuclear as
\begin{equation}
	\dot{M}^\mathrm{d}_\mathrm{nuc} = M_\mathrm{d}/\tau_\mathrm{nuc} = 1\times 10^{-10} (L_\mathrm{d}/L_\sun)\,M_\sun\,\mathrm{yr}^{-1}.
	\label{Mnuc}
\end{equation}
If the thermal (equation\,\ref{MKH}) or nuclear (equation\,\ref{Mnuc}) timescale mass transfer rate exceeds than the Eddington rate (equation\,\ref{Medd}), we assume the mass accretion rate is
\begin{equation}
\begin{aligned}
\dot{M}_\mathrm{acc} &= \eta_\mathrm{d} \dot{M}^\mathrm{d}_\mathrm{KH},~\mathrm{or}\\
                     &=\eta_\mathrm{d} \dot{M}^\mathrm{d}_\mathrm{nuc},
\end{aligned}
\end{equation}
with an efficiency $\eta_\mathrm{d}=0.1$. If the mass transfer rate is smaller than the Eddington limit, we assume $\eta_\mathrm{d}=1.0$, and
\begin{equation}
	\begin{aligned}
		L_\mathrm{acc} &=\eta_\mathrm{d} \eta_\mathrm{acc} \dot{M}^\mathrm{d}_\mathrm{KH} c^2, \mathrm{or} \\
		& =\eta_\mathrm{d} \eta_\mathrm{acc} \dot{M}^\mathrm{d}_\mathrm{nuc} c^2.
	\end{aligned}
	\label{Lacc}
\end{equation}

We use the observed IMXBs with available X-ray luminosity $L_\mathrm{X}$ and non-eccentric orbit to constrain the efficiency parameters $\eta_\mathrm{d}$ and $\eta_\mathrm{acc}$. Figure\,\ref{lx-imxb} shows that the upper tracks of the observed X-ray luminosity $L_\mathrm{X}$ are below $0.02 \dot{M}^\mathrm{d}_\mathrm{KH} c^2$. So the peak X-ray luminosity is described well by non-conserved ($\eta_\mathrm{d}=0.1$) thermal timescale mass transfer and super-Eddington accretion. The lower tracks of $L_\mathrm{X}$ are above $0.1 \dot{M}^\mathrm{d}_\mathrm{nuc} c^2$ (conserved $\eta_\mathrm{d}=1.0$ nuclear timescale mass transfer) or $0.01 \dot{M}^\mathrm{d}_\mathrm{nuc} c^2$ (non-conserved $\eta_\mathrm{d}=0.1$ nuclear timescale mass transfer and super-Eddington accretion).

The peak X-ray luminosity $L_\mathrm{X}$ of observed IMXBs spans over four orders of magnitudes. From the point of the energy contribution from the donor star, we find the peak X-ray luminosity can be explainded well by using thermal-(upper tracks) or nuclear-(lower tracks) timescale mass transfer. The upper tracks of $L_\mathrm{X}$ are derived from the non-coserved thermal timescale mass transfer, which is powered by a super-Eddington accretion. The lower tracks of $L_\mathrm{X}$ are calculated from nuclear timescale mass tranfer, which could be a conserved mass transfer ($M_\mathrm{d} < 4\,M_\sun$) or a non-conserved ($M_\mathrm{d} > 4\,M_\sun$) super-Eddington accretion. We simply assume the bolometric luminosity equals the X-ray luminosity. However, \citet{2021MNRAS.506.1045M} determine an intrinsic X-ray luminosity of $\ge 2\times 10^{37}{\rm \,erg\,s^{-1}}$ for SS 433. They infer that the hard X-ray emission from the inner regions is likely being scattered toward us by the walls of the wind-cone. If viewed face-on, they infer an apparent luminosity of $> 1\times 10^{39}{\rm \,erg\,s^{-1}}$. Furthermore, the optical/UV luminosity of SS 433 is in excess of $10^{40}{\rm \,erg\,s^{-1}}$ \citep{2019AA...624A.127W}. For super-Eddington accretion, it can be difficult to reliably relate $\dot{M}$ to $L_\mathrm{X}$ as the geometry of the accretion flow can introduce an isotropic in the radiation pattern. However, by observing changes in $P_\mathrm{orb}$ of M82 X-2, \citet{2022ApJ...937..125B} were able to place independent constraints on $\dot{M}$. This could allow us to avoid any issues with accretion efficiency or beaming.

\section{Summary}
\label{sec:sum}

This study is an extension of the series of Papers I, II and III which present systematically critical mass ratios for dynamical-timescale mass transfer over the span of donor star evolutionary states ($Z=0.02$). Using $4M_\sun$ donor stars as examples, we study the different responses of stars with metallicities $Z=0.02$ and $Z=0.001$, as well as their critical mass ratios. We present the critical mass ratios of IM stars with masses from $1.6$ to $10.0M_\sun$ with $Z=0.001$. Both a tabular form ($Z=0.001$ only) and a fitting formula ($Z=0.001$ and $Z=0.02$) of the critical mass ratios are provided in this paper. For metal-poor MS and HG donor stars, we find critical mass ratios are smaller than those of solar metallicity stars at the same evolutionary stage. However, for metal-poor RGB/AGB donor stars, we find critical mass ratios are larger than those of the solar metallicity stars with the same radii. Hence, metallicity has an important impact on the thresholds for dynamical-timescale mass transfer which leads to the common envelope evolution. We apply our results to 17 observed IMXBs with available mass ratios and orbital periods. We find our prediction constrains well on the observed IMXBs that undergo thermal or nuclear timescale mass transfer. We give a prediction of the upper and lower tracks to the X-ray luminosities of IMXBs as a function of the donor mass and the mass transfer timescale. This prediction based on the donor star might be a helpful complement to the accretion disk physics.

\section{acknowledgments}
We thank the anonymous referee for the constructive comments and suggestions on improving this paper. This project is supported by the National Key R\&D Program of China (2021YFA1600403) and the National Natural Science Foundation of China (grants NO. 12173081, 12090040/3, 11733008, 12125303), Yunnan Fundamental Research Projects (grant NO. 202101AV070001), the key research program of frontier sciences, CAS, No. ZDBS-LY-7005 and CAS, “Light of West China Program”. HG thanks the institute of astronomy, University of Cambridge, for hosting the one-year visit. HG also thanks Prof. Ronald Webbink for helpful discussions to build the adiabatic mass-loss model. CAT thanks Churchill College for his fellowship.

\bibliography{hongwei6}{}
\bibliographystyle{aasjournal}

\end{document}